\documentclass[12pt]{iopart}
\usepackage{siunitx}
\usepackage{color, graphicx,subcaption,url,booktabs}
\usepackage{tablefootnote}
\usepackage{csquotes}
\usepackage{bm}
\usepackage{harvard}[uniquelist=false,maxcitenames=1]
\citationmode{abbr}
\usepackage[colorlinks=true, linkcolor=blue, citecolor=blue, urlcolor=blue]{hyperref} 

\usepackage{textcomp,lineno,booktabs,amssymb} 
\begin{document}

\title[First experimental TOF-based proton radiography using LGADs]{First experimental time-of-flight-based proton radiography using low gain avalanche diodes}

\author{Felix Ulrich-Pur$^{1}$, Thomas Bergauer$^2$, Tetyana Galatyuk$^{1,3,4}$, Albert Hirtl$^5$, Matthias Kausel$^{5,6}$, Vadym Kedych$^3$, Mladen Kis$^1$, Yevhen Kozymka$^3$, Wilhelm Krüger$^3$, Sergey Linev$^1$, Jan Michel$^8$, Jerzy Pietraszko$^1$, Adrian Rost$^7$, Christian Joachim Schmidt$^1$, Michael Träger$^1$, Michael Traxler$^1$}

\address{$^1$GSI Helmholtzzentrum für Schwerionenforschung GmbH, Planckstraße 1, Darmstadt 64291, Germany}
\address{$^2$Austrian Academy of Sciences, Institute of High Energy Physics, Nikolsdorfergasse 18, Wien 1050, Austria}
\address{$^3$Technische Universität Darmstadt, Institut für Kernphysik, Karolinenplatz 5, Darmstadt 64289, Germany}
\address{$^4$Helmholtz Forschungsakademie Hessen für FAIR, Max-von-Laue-Straße 12, Frankfurt am Main 60438, Germany}
\address{$^5$TU Wien, Atominstitut, Stadionallee 2, Wien 1020, Austria}
\address{$^6$MedAustron, Marie-Curie-Straße 5, Wiener Neustadt 2700, Austria}
\address{$^7$FAIR Facility for Antiproton and Ion Research in Europe GmbH, Planckstraße 1, Darmstadt 64291, Germany}
\address{$^8$Goethe University Frankfurt, Institut für Kernphysik, Max-von-Laue-Str. 1, Frankfurt am Main 60438, Germany}
\ead{f.ulrich-pur@gsi.de}

\vspace{10pt}
\begin{indented}
\item[]December 2023
\end{indented}

\begin{abstract}
	\paragraph{Objective}	
	Ion computed tomography (iCT) is an imaging modality for the direct determination of the relative stopping power (RSP) distribution within a patient's body. Usually, this is done by estimating the path and energy loss of ions traversing the scanned volume utilising a tracking system and a separate residual energy detector. This study, on the other hand, introduces the first experimental study of a novel iCT approach based on time-of-flight (TOF) measurements, the so-called Sandwich TOF-iCT concept, which in contrast to any other iCT systems, does not require a residual energy detector for the RSP determination.
	\paragraph{Approach} A small Sandwich TOF-iCT demonstrator was built based on low gain avalanche diodes (LGADs), which are 4D-tracking detectors that allow to simultaneously measure the particle position and time-of-arrival with a precision better than $\SI{100}{\micro m}$ and $\SI{100}{ps}$, respectively. Using this demonstrator, the material and energy-dependent TOF was measured for several homogeneous PMMA slabs in order to calibrate the acquired TOF against the corresponding water equivalent thickness (WET). With this calibration, two proton radiographs (pRads) of a small aluminium stair phantom were recorded at MedAustron using $\SI{83}{MeV}$ and $\SI{100.4}{MeV}$ protons. 
	\paragraph{Main results} Due to the simplified WET calibration models used in this very first experimental study of this novel approach, the difference between the measured and theoretical WET ranged between $\SI{37.09}{\percent}$ and $\SI{51.12}{\percent}$. Nevertheless, the first TOF-based pRad was successfully recorded showing that LGADs are suitable detector candidates for Sandwich TOF-iCT.
	\paragraph{Significance} While the system parameters and WET estimation algorithms require further optimization, this work was an important first step to realize Sandwich TOF-iCT. Due to its compact and cost-efficient design, Sandwich TOF-iCT has the potential to make iCT more feasible and attractive for clinical application, which, eventually, could enhance the treatment planning quality.
\end{abstract}

\noindent{\it Time-of-flight, 4D-tracking, Low Gain Avalanche Diode, Ion Computed Tomography, Proton Therapy, Sandwich TOF-iCT\/}\\
\submitto{\PMB}
\section{Introduction}
In ion beam therapy, the quality of dose and range calculations strongly depends on precise knowledge of the tissue composition inside the treated volume. Especially the relative stopping power (RSP) of the irradiated tissue has to be precisely known as it describes the ion‘s energy loss per unit path length relative to the energy loss in water. Currently, RSP images are obtained via conventional x-ray computed tomography (CT), where the measured Hounsfield Units (HU) have to be converted to the corresponding RSP values \cite{Schneider1996}. This conversion, however, is a major source of inaccuracy, leading to RSP errors in the order of $\SI{1}{}$-$\SI{3}{\percent}$ \cite{Yang_2012} and $>\SI{0.6}{\percent}$ for more modern dual-energy CT (DECT) scanners \cite{Dedes2019}.\\
An alternative imaging modality that aims at improving the accuracy of the RSP estimation is ion computed tomography (iCT). Unlike CT or DECT, iCT allows determining the RSP directly by measuring the particle path and energy loss of single ions travelling through the patient using a tracking system and a separate calorimeter \cite{schulte_conceptual_2004}. While recent iCT prototype scanners showed promising results \cite{Johnson2017} and also the potential to compete with modern DECT scanners \cite{Dedes2019,bar_experimental_2022}, no clinical iCT system exists so far. The main reason is that meeting all detector requirements of a clinically viable iCT system is quite challenging, especially the minimum required data acquisition rate of at least $\SI{10}{MHz}$ \cite{Johnson2017} to keep the acquisition time of the iCT scan comparable to CT ($\mathcal{O}(\SI{}{min})$).\\
However, recent advances in 4D-tracking detectors, in particular the low gain avalanche diode (LGAD) technology, constitute a potential solution to this challenge. LGADs are silicon-based particle detectors, which allow to simultaneously measure the particle's position and time-of-arrival (ToA) with a spatial and time precision better than $\SI{100}{\micro m}$ and $\SI{100}{ps}$, respectively \cite{Sadrozinski2017}. Consequently, single-particle tracks can be resolved even at very high track densities, thus leading to much higher rate capabilities when compared to conventional silicon detectors. For instance, in a recent proton-proton production experiment with the high-acceptance dielectron spectrometer (HADES) at GSI, an LGAD-based start reaction time detector and beam monitor system was able to cope with a particle flux of $\SI{e8}{p/s/cm^2}$ \cite{kruger_lgad_2022}. Besides this high rate performance, LGADs are also very radiation hard as they can withstand neutron equivalent fluences of up to $\approx\SI{e15}{}\mathrm{n}_\mathrm{eq}/\SI{}{cm^2}$ without significantly deteriorating their excellent 4D-tracking capability \cite{Padilla2020}. All those detector properties, combined with a relatively low material budget ($\mathcal{O}(\SI{100}{\micro m}$) silicon), make LGADs a perfect detector candidate for high luminosity environments with many applications ranging from high energy physics to medical physics, e.g. iCT.\\
Building an iCT system based on LGADs could not only help to boost the scanner's rate capability, but it would also allow integrating time-of-flight (TOF) measurements into the imaging process itself, which is then referred to as TOF-iCT. In contrast to conventional iCT designs, a TOF-iCT system can be solely based on LGADs, as LGADs can be used for both the particle path and energy loss estimation. For instance, by measuring the TOF in air inside a TOF calorimeter placed downstream of the patient, the residual energy of the ion can be determined \cite{volzphd2021,krah_relative_2022,TOFcalUlrich-Pur_2022}. Alternatively, the energy and material-dependent increase in an ion's TOF through the scanned object can be used as an indirect measure for the energy loss, which is the so-called Sandwich TOF-iCT concept introduced in \citeasnoun{sandwichTOFUlrich-Pur_2023}. The latter approach, unlike any other iCT system, does not require a dedicated residual energy detector downstream of the patient, which would make the scanner design much more compact and therefore easier to integrate into a clinical treatment room.\\
While recent Monte Carlo (MC) feasibility studies demonstrated the potential of both TOF-iCT \cite{TOFcalUlrich-Pur_2022,krah_relative_2022} and Sandwich TOF-iCT \cite{sandwichTOFUlrich-Pur_2023} to fulfil the requirements of a clinical iCT system, no TOF-iCT system exists so far. This study aims to fill this gap by providing the first experimental study of an LGAD-based iCT system, focussing on the novel Sandwich TOF-iCT method. This should serve both as an experimental proof-of-concept for Sandwich TOF-iCT as well as a guide for potential refinements in succeeding TOF-iCT designs and methods. For that purpose, we designed a compact Sandwich TOF-iCT prototype using single-sided LGAD strip sensors. The LGAD sensors were obtained from an R\&D sensor production run at the Fondazione Bruno Kessler (FBK) and operated using a modified readout system from the LGAD-based start reaction time detector of the HADES experiment at GSI \cite{kruger_lgad_2022}.\\ Utilising this demonstrator, we generated a proton radiography (pRad) of a small aluminium stair phantom at the research and therapy centre MedAustron in Wiener Neustadt, Austria. Details about the calibration procedures of the Sandwich TOF-iCT system, encompassing both sensor and water-equivalent-thickness (WET) calibration, will be outlined in the following before delving into the specifics of the pRad experiment, its image reconstruction and analysis.

\section{Material and Methods}
\subsection{Sandwich TOF-iCT concept}
The Sandwich TOF-iCT concept was first introduced in \citeasnoun{sandwichTOFUlrich-Pur_2023} and will be summarized briefly in the following.
\subsubsection{Time-of-flight of ions in matter}\mbox{}\\
As ions travel through matter, they lose energy and slow down, which leads to a material and energy-dependent TOF, calculated as 
\begin{equation}
    \mathrm{TOF}=\int\limits_0^L \frac{\mathrm{d}s}{v(\vec{\mathbf{x}}(s))}=\int\limits_0^L \frac{\mathrm{d}s}{c \cdot \frac{E_\mathrm{kin}(\vec{\mathbf{x}}(s))}{E_\mathrm{kin}(\vec{\mathbf{x}}(s))+m_0c^2}\sqrt{1+2\cdot \frac{m_0c^2}{E_\mathrm{kin}(\vec{\mathbf{x}}(s))}}}
    \label{eq:tof}
\end{equation}
with $m_0$ as the rest mass of the ion, $c$ as the speed of light and $E_\mathrm{kin}(\vec{\mathbf{x}}(s))$ and $v(\vec{\mathbf{x}}(s))$ as the kinetic energy and corresponding velocity at position $\vec{\mathbf{x}}(s)$. 
\subsubsection{Slowing Down Power}\mbox{}\\
While the TOF in equation (\ref{eq:tof}) strongly depends on the tissue composition, beam energy and ion's path, it was shown that, for a small pathlength increment $\Delta x$ inside the traversed tissue, the difference in TOF per unit pathlength with respect to the TOF in vacuum ($\mathrm{TOF}_\mathrm{vac}$), i.e. without any energy loss,  is directly related to the stopping power (SP) of the medium according to 
\begin{equation}
    \frac{\mathrm{TOF}-\mathrm{TOF}_\mathrm{vac}}{\Delta x}=\frac{\Delta \mathrm{TOF}}{\Delta x}:= \mathrm{SDP}(E) = f(E) \cdot \mathrm{SP} (E)
    \label{eq:sdp}
\end{equation}
with $f(E)$ as a solely energy-dependent term, $\Delta \mathrm{TOF}=\mathrm{TOF}-\mathrm{TOF}_\mathrm{vac}$ and $\mathrm{SDP}(E)$ as the so-called slowing-down power of the traversed medium at the beam energy $E$.
\subsubsection{Relative Slowing Down Power}\mbox{}\\
Using equation (\ref{eq:sdp}), one can easily see that the relative slowing down power (RSDP), i.e. the SDP in matter ($\mathrm{SDP}_\mathrm{mat}$) relative to the SDP in water ($\mathrm{SDP}_{\mathrm{H}_2\mathrm{O}}$) is equal to the RSP since $f(E)$ cancels out and 
\begin{equation}
    \mathrm{RSDP}:= \frac{\mathrm{SDP}_\mathrm{mat}(E)}{\mathrm{SDP}_{\mathrm{H}_2\mathrm{O}}(E)} =\frac{\mathrm{SP}_\mathrm{mat}(E)}{\mathrm{SP}_{\mathrm{H}_2\mathrm{O}}(E)} = \mathrm{RSP}
    \label{eq:rsdp}
\end{equation}
with $\mathrm{SP}_\mathrm{mat}$ and $\mathrm{SP}_{\mathrm{H}_2\mathrm{O}}$ denoting the SP in the material and water, respectively.
\subsubsection{Imaging problem}\mbox{}\\
In standard iCT, the RSP distribution inside the patient can be reconstructed by measuring the particle path of multiple ions travelling through the patient and the corresponding WETs, which are a measure for the total energy loss along the path ($\Delta E$) for a given primary energy $E_0$. The respective inverse problem reads as follows
\begin{equation}
   \mathrm{WET}(E_0,\Delta E)=\int\limits_0^L \mathrm{RSP}(\vec{\mathbf{x}}(s))\mathrm{d}s \approx \int\limits_0^L \mathrm{RSDP}(\vec{\mathbf{x}}(s))\mathrm{d}s,
   \label{eq:inverseprob}
\end{equation}
with $\mathrm{RSP}(\vec{\mathbf{x}}(s))$ and $\mathrm{RSDP}(\vec{\mathbf{x}}(s))$ denoting the RSP and RSDP at positon $\vec{\mathbf{x}}(s)$ and $L$ the total pathlength of the ion's trajectory inside the patient. Inserting equations (\ref{eq:sdp}) and (\ref{eq:rsdp}) into the right-hand-side of equation (\ref{eq:inverseprob}) yields the following image reconstruction problem for Sandwich TOF-iCT 
\begin{equation}
    \int\limits_0^{\mathrm{TOF}-\mathrm{TOF}_\mathrm{vac}}\frac{\mathrm{d}\Delta \mathrm{TOF}}{\mathrm{SDP}_{\mathrm{H}_2\mathrm{O}}(\Delta \mathrm{TOF}(E(\vec{\mathbf{x}}(s))))}=\int\limits_0^L \mathrm{RSP}(\vec{\mathbf{x}}(s))\mathrm{d}s,
    \label{eq:imagerecosandwich}
\end{equation}
where $\mathrm{SDP}_{\mathrm{H}_2\mathrm{O}}$ is given as a function of $\Delta \mathrm{TOF}(E(\vec{\mathbf{x}}(s)))$. To obtain the relation between $\Delta \mathrm{TOF}$ and $E$, and therefore $\mathrm{SDP}_{\mathrm{H}_2\mathrm{O}}(\Delta \mathrm{TOF}(E(\vec{\mathbf{x}}(s))))$, one can correlate the cumulative TOF increase in water for different water thicknesses and a given primary beam energy $E_0$ to the corresponding residual energy $E(\vec{\mathbf{x}}(s))$ using the initial condition $\Delta \mathrm{TOF}(E(\vec{\mathbf{x}}(s=0)))=0$. The correlation $\Delta \mathrm{TOF}(E(\vec{\mathbf{x}}(s)))$ for water can be either obtained via MC simulations or analytically as described in \citeasnoun{sandwichTOFUlrich-Pur_2023}. After determining $\mathrm{SDP}_{\mathrm{H}_2\mathrm{O}}(\Delta \mathrm{TOF}(E(\vec{\mathbf{x}}(s))))$ and measuring the total increase in TOF ($\mathrm{TOF}-\mathrm{TOF}_\mathrm{vac}$) and the total pathlength $L$, one can reconstruct the RSP using standard iCT reconstruction algorithms, e.g. as defined in \citeasnoun{Rit2013}. 

\subsubsection{Alternative WET calibration approach}\mbox{}\\
\label{sec:wetcalibmeth}However, actually measuring the total increase in TOF requires prior knowledge of the velocity distribution along the particle path, which can only be estimated, e.g. via MC simulation. Therefore, as a first step, a much simpler approach was introduced in \citeasnoun{sandwichTOFUlrich-Pur_2023}. Instead of estimating the true increase in TOF, the increase in TOF w.r.t the TOF in air ($\mathrm{TOF}_\mathrm{air}$), i.e. the TOF through the scanner without the phantom, is determined. For each particle, the resulting TOF increase is then mapped to the corresponding WET of the traversed medium using a fifth-order polynomial
\begin{equation}
    \mathrm{TOF}-\mathrm{TOF}_\mathrm{air} (\mathrm{WET}, E_0 ) = \sum_{i=0}^5 a_i(E_0) \cdot \mathrm{WET}^i
    \label{eq:WETcalibmodel}
    \label{eq:wetcalib}
\end{equation}
with $a_i(E_0)$ as the fit parameters for an ion with primary beam energy $E_0$. To obtain the fit parameters $a_i(E_0)$, a calibration run has to be performed prior to the actual imaging experiment, where $\mathrm{TOF}-\mathrm{TOF}_\mathrm{air} (\mathrm{WET}, E_0 )$ has to be measured for a fixed beam energy $E_0$ and different absorbers with known WET.
\\\\
While this simplified calibration model is easy to implement, it also introduces a systematic dependence of the WET estimation on the system parameters of the used TOF-iCT system \cite{sandwichTOFUlrich-Pur_2023}. Thus, an optimized model should be developed to further advance this imaging modality. However, since the study presented here focuses mainly on the first experimental realisation of Sandwich TOF-iCT, improving the WET estimation algorithms for this modality was out of the scope. Therefore, the simplified WET calibration method described in equation (\ref{eq:wetcalib}) was also chosen for the pRad experiment conducted at MedAustron.

\subsection{TOF-iCT demonstrator}
The TOF-iCT demonstrator utilizes an adapted version of the readout system for the LGAD-based HADES start-reaction time ($\mathrm{T}_0$) detector developed at GSI \cite{kruger_lgad_2022}. A brief description of all components of the TOF-iCT demonstrator, including the LGAD sensors, corresponding front-end electronics (FEE) and data-acquisition (DAQ) system, is given in the following. 
\subsubsection{LGAD sensors}
\begin{figure}[ht]
	\centering
 \begin{subfigure}[b]{0.48\textwidth}
		\centering
		\includegraphics[height=6cm]{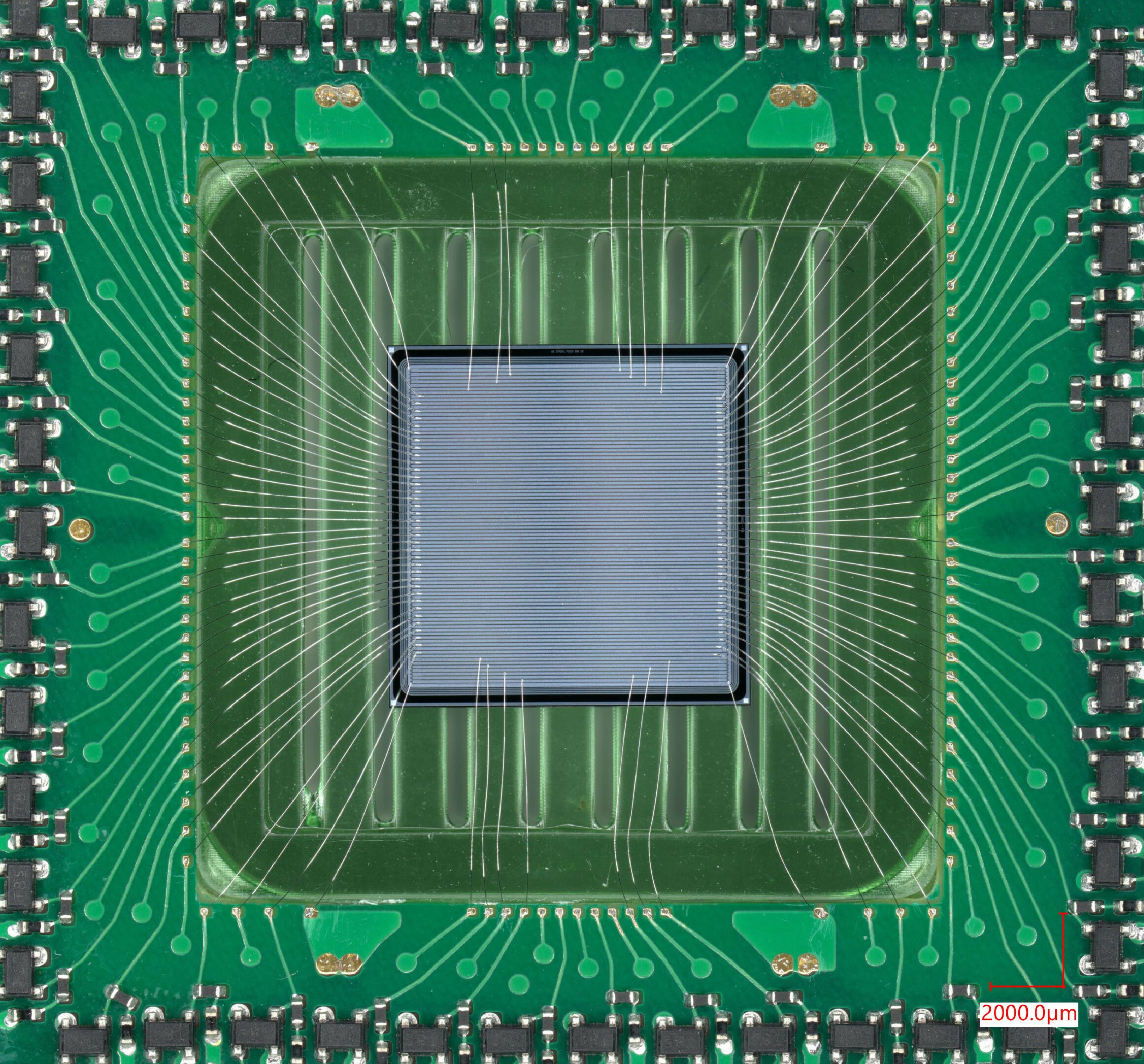}
		\caption{$\SI{1}{}\times \SI{1}{cm^2}$ LGAD strip sensor with 86 channels wire-bonded to the discrete front-end electronics.}
		\label{fig:lgadsensor}
	\end{subfigure}
 \quad
 \begin{subfigure}[b]{0.48\textwidth}
		\centering
		\includegraphics[height=6cm]{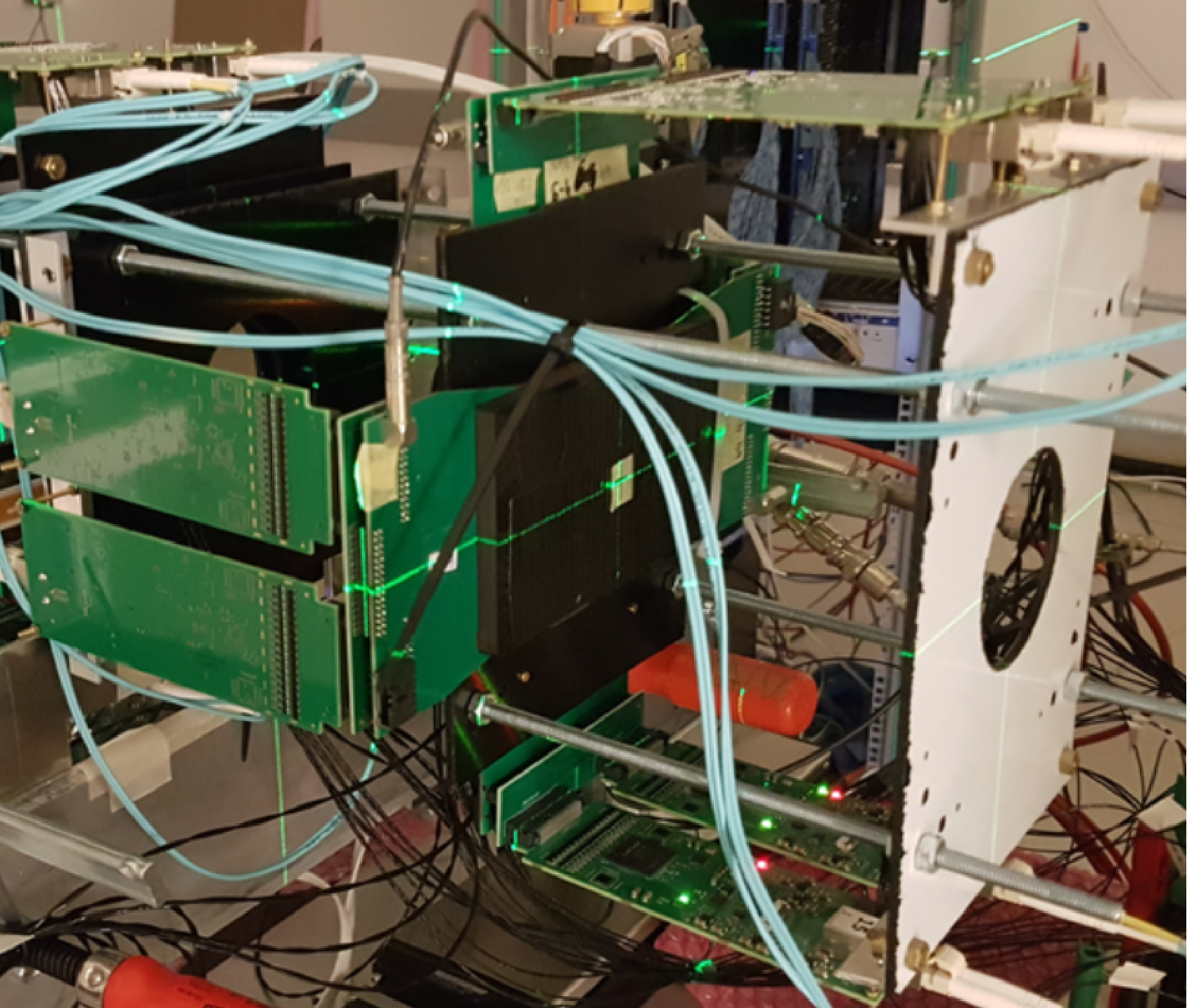}
		\caption{Two single-sided LGAD strip sensor modules were mounted perpendicular to each other to form a full 4D-tracking layer.}
		\label{fig:lgadxymodule}
	\end{subfigure}
	\begin{subfigure}[b]{0.99\textwidth}
		\centering
		\includegraphics[height=5cm]{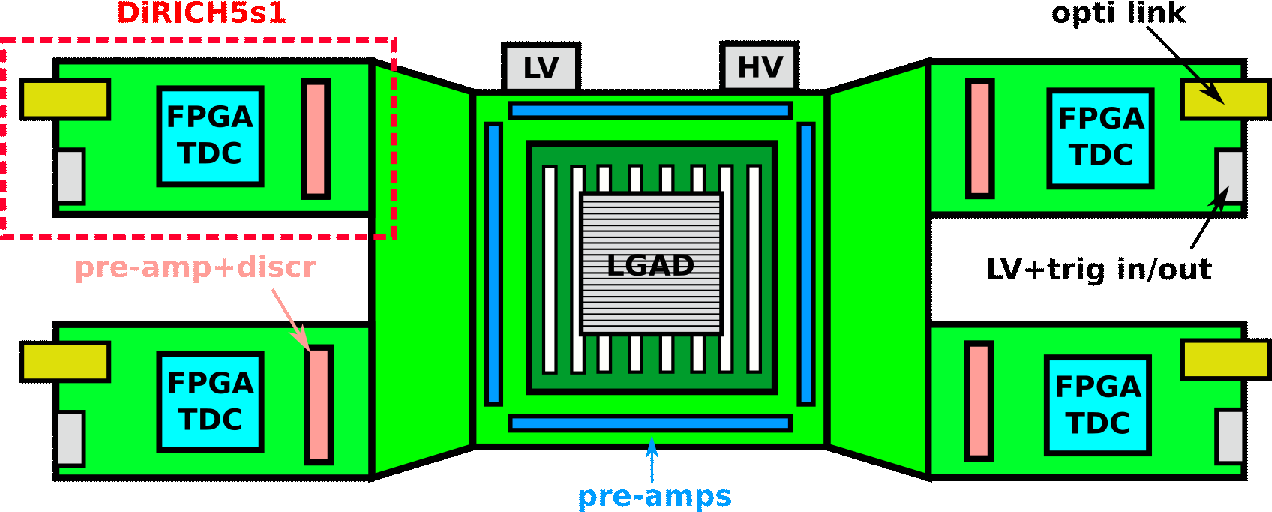}
		\caption{Schematic of a single-sided LGAD strip sensor module. } 
		\label{fig:lgadmodule}
	\end{subfigure}\vspace{0.3cm}

	\caption{Components of the LGAD-based 4D-tracking module.}
\end{figure}

\mbox{}\\
Four single-sided LGAD strip sensors were used for the TOF-iCT demonstrator, each measuring $\approx \SI{1}{}\times \SI{1}{cm^2}$ and containing 86 strips with a strip length of $\SI{8.6}{mm}$ and a pitch of $\SI{100}{\micro m}$ (figure \ref{fig:lgadsensor}). Similar to the  HADES $\mathrm{T}_0$ detector, which stems from the same  R\&D production run as the LGADs used within this work, all sensors underwent a thinning procedure to reduce the overall material budget of the detector, resulting in a total sensor thickness of $\SI{200}{\micro m}$. 
\subsubsection{LGAD module}\mbox{}\\
In order to bias and read out each sensor, four detector modules were developed. As shown in figure \ref{fig:lgadmodule}, each module was designed with a centrally located rectangular cut-out in which a 3D-printed plastic grid was placed (dark green grid). The sensor was then glued onto this grid and connected to the discrete FEE consisting of a two-stage pre-amplifier indicated as dark blue strips and the corresponding low-voltage (LV) power lines to drive the pre-amplifiers as well as high-voltage (HV) power lines for biasing the LGAD. The pre-amplified signals were then forwarded to custom FPGA-based time-to-digital converter (TDC) boards (DiRICH5s1 boards), which also feature another amplification stage and a leading edge discriminator (indicated as pink strips) as well as an optical link for the readout (highlighted in yellow) and low voltage differential signal (LVDS) lines for triggering (grey connectors). Using the DiRICH5s1, the leading and falling edge of the amplified LGAD signal could be determined, which allowed measuring the time-of-arrival (ToA) of the particle as well as the time-over-threshold (TOT), i.e. a measure of the signal duration and therefore also signal height. While each DiRICH5s1 has 32 readout channels, the channel configuration inside the discrete FEE required four of those FPGA-based TDC boards to read out all 86 channels of a single LGAD sensor. However, constraints related to the number of available DiRICH5s1 boards, which was 14 in total, meant that only every second channel in the final LGAD was effectively connected. This still allowed full particle tracking inside the last sensor as the cluster size, i.e. the number of fired strips per particle, was mostly greater than 3 for the used proton beam energies, and therefore, this sensor required a minor adaptation of the 4D-cluster finder algorithm, which will be briefly outlined in section \ref{sec:4dclustgmeth}. 
\subsubsection{4D-tracking layer}\mbox{}\\
Given that an individual LGAD module can measure just one spatial coordinate, we paired two of the aforementioned modules orthogonally to each other to create a single 4D-tracking layer. Since mechanical constraints prohibited mounting the LGADs in close proximity, they were installed at a distance of $\SI{1.64}{cm}$ between the individual LGAD sensors. For each sensor, an additional light-tight enclosure was constructed using a 3D-printed plastic holder featuring a circular aperture with an area greater than the LGAD size. This cutout was subsequently sealed with light-tight tape to ensure total light isolation. An image of a completely assembled 4D-tracking layer can be seen in figure \ref{fig:lgadxymodule}. 
\begin{figure}[h]
	\centering
	\includegraphics[height=5cm]{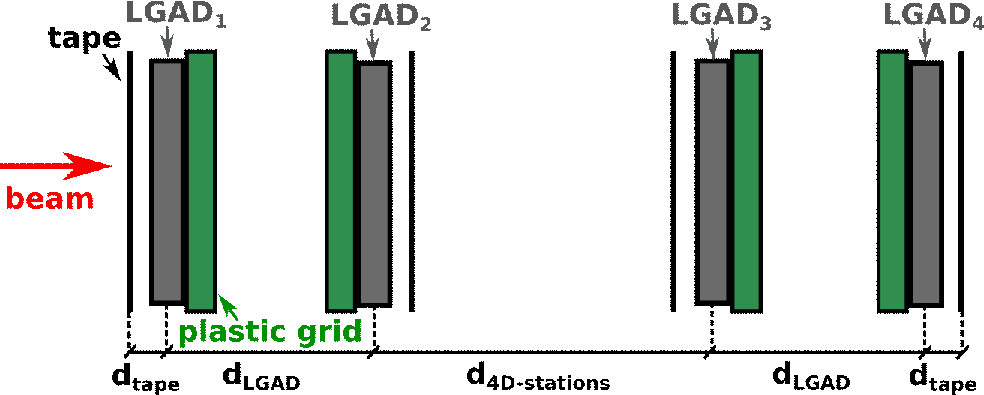}
		\caption{Schematic representation of the TOF-iCT demonstrator setup at MedAustron, highlighting only components directly irradiated by the beam and positioned in the field of view of the LGAD sensors, thus influencing the experimental outcome.} 
		\label{fig:demonstratorlayout}
\end{figure}

\subsubsection{TOF-iCT demonstrator layout}\mbox{}\\
Figure \ref{fig:demonstratorlayout} illustrates the structural design of the TOF-iCT demonstrator, which consists of two 4D-tracking layers separated by a distance of $\mathrm{d}_\mathrm{4D-stations}=\SI{27}{cm}$. While $\mathrm{d}_\mathrm{LGAD}$ denotes the previously mentioned spacing between the individual LGAD sensors per 4D-tracking layer, $\mathrm{d}_\mathrm{tape}$ describes the gap between the sensor and the light-tight tape. The latter was determined by the dimensions of the light-tight enclosure, measuring $\SI{0.5}{cm}$.
\begin{figure}[b]
    \centering
		\includegraphics[width=0.99\textwidth]{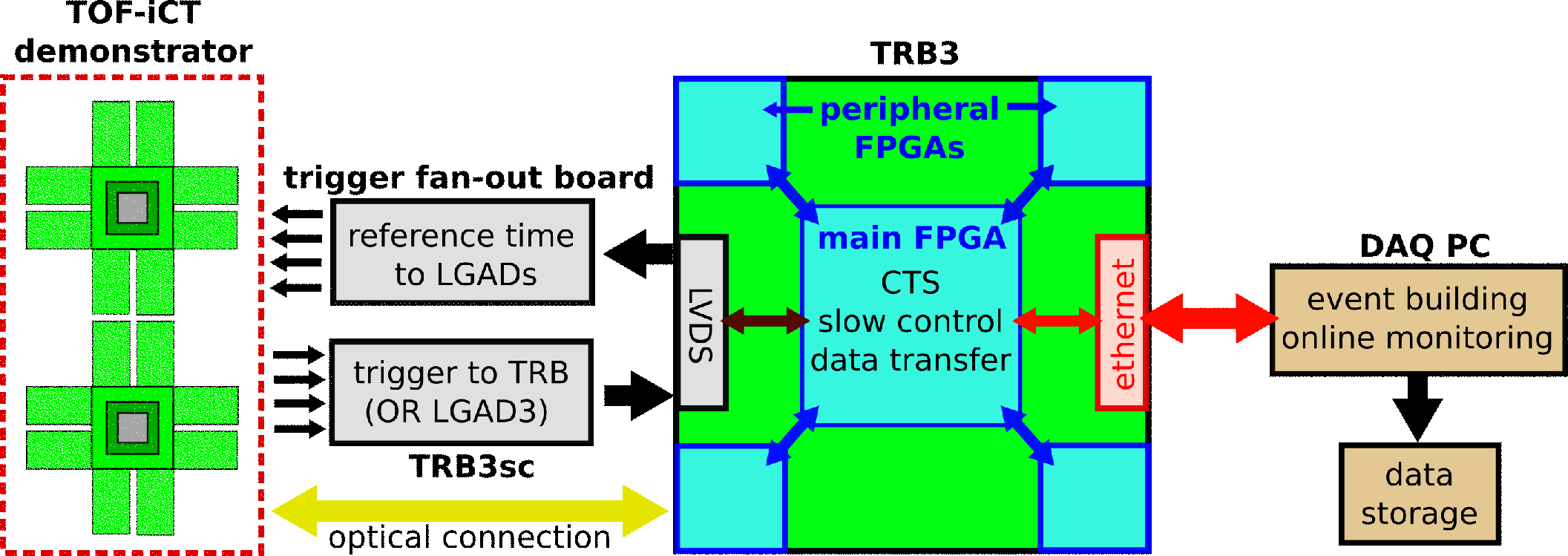}
		\caption{Sketch of the TOF-iCT demonstrator DAQ system.}
		\label{fig:daqlayout}
\end{figure}
\subsubsection{Data acquisition system}\mbox{}\\
The DAQ system of the TOF-iCT demonstrator (figure \ref{fig:daqlayout}) is a modified version of the HADES DAQ system \cite{michel_upgraded_2011}. Central to this system is the TrbNet \cite{michel_-beam_2013}, a custom FPGA-based network protocol used for triggering, data transfer and slow control of the FEE of the individual detectors, e.g. the DiRICH5s1 boards on the LGAD modules. The main hardware component of the DAQ system is the FPGA-based multi-purpose trigger and readout board version 3 (TRB3), which houses five FPGAs \cite{neiser_trb3_2013}. One of these FPGAs acts as the central controller and the other four establish the optical connections to the LGAD modules using supplementary adapter boards equipped with small form-factor pluggable (SFP) modules. While the synchronisation, data transfer and slow control operations are carried out through these optical links, additional LVDS connections provide the physical trigger to the DiRICH5s1 boards. Both the synchronisation and triggering are managed by the so-called central triggering system (CTS), which is also integrated into the central controller FPGA. Leveraging the capability of the CTS to accept external trigger inputs, an auxiliary FPGA-based logic board (i.e. the Trb3sc) was utilized to set the trigger condition for the readout. A logical OR across all channels of the third LGAD was chosen for this purpose, meaning that a trigger was initiated and forwarded to the CTS inside the TRB3 via LVDS whenever the signal in any channel of the third LGAD exceeded the threshold set in the corresponding leading-edge discriminators. The CTS then generated an output trigger, which was synchronously dispatched to all the LGAD modules through a fan-out board linked to the TRB3. Once an event was triggered, all incoming signals that fell within a predefined time window were sent from the DiRICH5s1 boards to the main FPGA and then to the readout PC via ethernet. For this experiment, the time window was set to $\SI{250}{ns}$ prior to and $\SI{50}{ns}$ after the trigger's arrival to account for the different delays in the readout chain and to guarantee the simultaneous measurement of a single particle in all detector modules. On the readout PC, the C++-based data acquisition backbone core framework (DABC) and Go4 analysis framework (Go4) were running to combine the data streams and build complete events, perform online monitoring and store the pre-processed data to disk. A more detailed description of the HADES DAQ and all used components and protocols can be found on the \citename{TRBHomepage} (\href{trb.gsi.de}{trb.gsi.de}).
\subsection{Sensor calibration}\label{sec:sensorcalibmeth}
Prior to recording the pRad, the LGAD sensors had to be calibrated. The calibration procedures, which will be outlined in the following sections, were first tested and optimized in the detector laboratory at GSI using a Sr-90 source. Given that the Sr-90 electrons produce signals similar to minimum-ionizing particles (MIPs), it was necessary to repeat the calibration for each of the used proton beam energies at MedAustron to cover the full signal height spectrum. Since the used beam energies varied between $\SI{83}{MeV}$ and $\SI{800}{MeV}$, the signal heights were anticipated to exceed that of MIPs by a factor ranging between $1.14$ and  $4.03$.
\subsubsection{TDC calibration and threshold scan}\mbox{}\\
First, each TDC had to be calibrated using internal calibration pulses issued from the CTS. Details about the FPGA-based TDC calibration can be found on the \citename{TRBHomepage} (\href{trb.gsi.de}{trb.gsi.de}). Then, the thresholds in each leading-edge discriminator were set to a fixed value above the noise level ($\approx \SI{20}{mV}$), which was low enough to capture the smaller, capacitively-coupled signals in the strips adjacent to the actual strip hit by the particle. 
Since the noise level was assumed to be constant for the entire experiment, the thresholds were only set at the beginning of the experiment.
\subsubsection{Time-over-threshold re-scaling}\label{sec:totnormmeth}\mbox{}\\
However, slight variations in threshold levels, attributed to different noise levels, and differences in signal amplification across individual LGAD channels could still be expected. Given that these factors can significantly impact the signal response, especially the obtained ToT, normalizing the ToT measurement for each channel was essential to maintain a uniform response across all sensors. Therefore, each 4D-tracking module was irradiated with protons of a fixed beam energy $E_0$ and the corresponding ToT distributions were recorded in both sensors of every 4D-tracking module. First, the most probable ToT value ($\mathrm{TOT}_{\mathrm{MPV}}|_{E_0}$) was determined in each LGAD channel by fitting a Gaussian to the signal peak of the obtained ToT distributions. Then, the ToT spectra were normalized by setting the $\mathrm{TOT}_{\mathrm{MPV}}|_{E_0}$ to $\SI{20}{ns}$ in every LGAD channel using the following re-scaling function  
\begin{equation}
    \label{eq:totrescale}
    \mathrm{TOT}(i,j)=\frac{\SI{20}{ns}}{\mathrm{TOT}_{\mathrm{MPV}}|_{E_0}(i,j)}*\mathrm{TOT}_\mathrm{raw}(i,j).
\end{equation}
Here, $\mathrm{TOT}_\mathrm{raw}(i,j)$ and $\mathrm{TOT}(i,j)$ denote the raw and the corresponding normalized ToT value measured in channel $i$ of LGAD $j$.

\subsubsection{Time-walk and offset correction}\label{sec:timewalkmeth}\mbox{}\\
Since both the ToA and the ToT are determined using a predefined, fixed threshold, their measured values strongly depend on the actual amplitude of the physical signal. This amplitude-dependence of the ToA, also known as the time-walk effect, can lead to ToA variations in the order of several nanoseconds, depending on the used readout electronics and resulting signal shape. As those variations can significantly deteriorate the precision of the time measurement, correcting for the time-walk effect is essential. The time-walk correction in each 4D-tracking layer was done in two steps using the data sets recorded for the ToT normalisation (section \ref{sec:totnormmeth}). First, to obtain the time-walk trend, the time difference spectra between every channel on one LGAD and a reference channel on the other LGAD of a single 4D-tracking layer were determined with respect to the corresponding signal amplitudes on the first LGAD, given via the related ToT of those signals. A 4D-clustering procedure (section \ref{sec:4dclustgmeth}) in addition to a ToT cut to the signals in the reference channels was applied to avoid redundancy by selecting only hits, which stem from the actual particle and not from noise events or capacitive-coupling between the neighbouring strips. Then, to estimate the time walk trend in each LGAD strip, the corresponding time difference vs ToT spectrum was divided into $\SI{50}{ps}\times \SI{50}{ps}$ bins and the most probable time difference value ($\mathrm{T}_\mathrm{Diff,MPV}(\mathrm{ToT})$) was obtained for each ToT bin using a Gaussian fit. The obtained $\mathrm{T}_\mathrm{Diff,MPV}(\mathrm{ToT})$ values were then stored in a look-up-table (LUT) and used to remove the signal amplitude dependence of the measured ToA in this channel. This was done by determining the $\mathrm{T}_\mathrm{Diff,MPV}(\mathrm{ToT})$ for the measured ToT and subtracting it from the corresponding ToA ($\mathrm{ToA}_{\mathrm{meas}}(\mathrm{ToT})$) according to 
\begin{equation}
    \label{eq:timealk}
    \mathrm{ToA}_\mathrm{corr}=\mathrm{ToA}_{\mathrm{meas}}(\mathrm{ToT})-\mathrm{T}_\mathrm{Diff,MPV}(\mathrm{ToT}),
\end{equation}
with $\mathrm{ToA}_\mathrm{corr}$ as the time-walk corrected ToA. After calibrating every channel inside the first LGAD, the same procedure was applied to the second LGAD of the same 4D-tracking layer by choosing one channel in the first LGAD as the reference channel. \\\\
While the previously mentioned time-walk calibration procedure removes any ToT dependence of the measured ToA, it also has the advantageous effect of synchronising the time measurement across the entire 4D-tracking layer. By subtracting $\mathrm{T}_\mathrm{Diff,MPV}(\mathrm{ToT})$ from the measured ToA, the mean time difference between the calibrated channel on the one LGAD and corresponding reference channel on the second LGAD of the same 4D-tracking layer will always be zero for the same beam energy. Since the same reference channel on one LGAD is used to calculate $\mathrm{T}_\mathrm{Diff,MPV}(\mathrm{ToT})$ for calibrating all channels on its partner LGAD, the mean time difference between all those channels and the corresponding reference channel is also zero. This also means that no offset in the measured ToA between the individual LGAD channels caused e.g. different signal propagation times inside the individual channels or the non-zero propagation time of the particles travelling between the sensors should remain. Consequently, for the same beam energy, all channels inside the 4D-tracking layer should measure precisely the same mean ToA with the ToA of the reference channel defining the global reference time. However, since the time-walk calibration procedure has to be done once for one LGAD and once for the partnering LGAD of the same 4D-tracking layer, it is important to carefully choose the order of the calibration as this will define the final reference ToA for the entire 4D-tracking layer. For instance, if $\mathrm{LGAD}_\mathrm{1}$ (figure \ref{fig:demonstratorlayout}) is calibrated after $\mathrm{LGAD}_\mathrm{2}$, the reference channel in $\mathrm{LGAD}_\mathrm{1}$ defines the final reference ToA of the corresponding 4D-tracking layer. For all following experiments, one central strip in $\mathrm{LGAD}_\mathrm{2}$ and one central strip $\mathrm{LGAD}_\mathrm{3}$ were used to define the final reference ToAs.

\subsection{Hit reconstruction and 4D-tracking}\label{sec:4dtrackandclustmeth}
\subsubsection{Hit reconstruction}\label{sec:4dclustgmeth}\mbox{}\\
As mentioned in section \ref{sec:timewalkmeth}, a particle hit inside the LGADs results in a signal above the discriminator threshold across a cluster of strips due to capacitive coupling between the individual detector channels. While the strip closest to the actual particle hit position yields the highest signal (highest ToT), the neighbouring strips detect a reduced signal, depending on the capacitive properties of the LGAD strip detector. To estimate the actual hit position for each particle, those clusters have to be identified and analyzed. For that purpose, a 4D cluster finder algorithm was implemented, which will be summarized briefly in the following. For every hit inside the LGAD, coincident signals within a given time window ($\pm \SI{15}{ns}$ before and $\pm \SI{1}{ns}$ after the time-walk calibration) and strip distance to the strip position of the corresponding hit ($\pm \SI{12}{}$ strips) were determined. For $\mathrm{LGAD}_\mathrm{1}$, $\mathrm{LGAD}_\mathrm{2}$ and $\mathrm{LGAD}_\mathrm{3}$ (figure \ref{fig:demonstratorlayout}), where all channels could be read out, a cluster was found if the strips of those coincident hits were consecutive in space. For the last LGAD ($\mathrm{LGAD}_\mathrm{4}$), where only every second channel was connected, a cluster was found if the coincident hits where within the previously described time window and strip range. Once a 4D cluster was found, the true ToA and hit position of the particle hit inside the LGAD were estimated using the strip signal inside the cluster with the largest ToT, i.e. closest to the actual particle position. 
\subsubsection{4D-track selection}\label{sec:4dtrackingmeth}\mbox{}\\
After applying the 4D cluster finder procedure to the calibrated data and estimating the corresponding particle hit positions and ToAs for each cluster, the 4D-particle tracks inside the TOF-iCT demonstrator could be reconstructed using a custom 4D-tracking algorithm. However, given that the TOF-iCT demonstrator consists of two 4D-tracking layers, only a straight-line track model could be utilised. First, a track candidate was identified if every LGAD had at least one cluster within a triggered event. Then, to simplify the 4D-tracking finding procedure, only events with exactly one cluster per layer were selected for the subsequent analysis to guarantee that only one particle passed through the TOF-iCT demonstrator.
\subsubsection{Detector alignment and position measurement}\label{sec:alignmentmeth}\mbox{}\\
In order to align the 4D-tracking layers of the TOF-iCT demonstrator, an available laser-positioning system at MedAustron was utilised. Following this, a more accurate, software-based detector alignment technique similar to the pre-alignment procedure defined in \citeasnoun{dannheim_corryvreckan_2021} was applied. For that purpose, the TOF-iCT demonstrator was irradiated with $\SI{800}{MeV}$ protons and the hit positions were recorded on both 4D-tracking layers using the 4D-track candidates described in section {\ref{sec:4dtrackingmeth}}. Then, the lateral displacement $r_x$ and $r_y$ between the first and second 4D-tracking layer were calculated for each particle according to
\begin{equation}
    r_x=x_\mathrm{back}-x_\mathrm{front} \quad \text{and} \quad r_y=y_\mathrm{back}-y_\mathrm{front},
\end{equation}
with $x_\mathrm{front}$, $y_\mathrm{front}$, $x_\mathrm{back}$, $y_\mathrm{back}$ denoting the $x$ and $y$ hit position in the first and second 4D-tracking layer, respectively. Any lateral misalignment of the layers would then manifest as a non-zero mean in the distributions of $r_x$ and $r_y$ since the average trajectory of all particles should theoretically be a straight line. To mitigate this misalignment, the first 4D-tracking layer was designated as the reference layer and the second layer's hit positions were adjusted by subtracting the mean offsets $\hat{r}_x$ and $\hat{r}_y$, which were obtained by fitting a 1D Gaussian to both the $r_x$ and $r_y$ distributions.
\subsubsection{TOF measurement}\mbox{}\\
Using the calibrated and aligned TOF-iCT demonstrator, the TOF through the scanner could be determined. For all of the following experiments, the TOF was defined as the difference between the mean ToA per 4D-tracking layer using
\begin{equation}
    \label{eq:tofinscanner}
    \mathrm{TOF}=\frac{\mathrm{TOA}_\mathrm{corr}(3)+\mathrm{TOA}_\mathrm{corr}(4)}{2}-\frac{\mathrm{TOA}_\mathrm{corr}(1)+\mathrm{TOA}_\mathrm{corr}(2)}{2}
\end{equation}
with $\mathrm{TOA}_\mathrm{corr}(j)$ as the calibrated ToA measured in LGAD $j$. As detailed in section \ref{sec:sensorcalibmeth}, the reference ToA channels for the offsets correction between the individual LGAD layers were chosen such, that the measured TOF in equation (\ref{eq:tofinscanner}) reflects the TOF between $\mathrm{LGAD}_\mathrm{2}$ and $\mathrm{LGAD}_\mathrm{3}$ (figure \ref{fig:demonstratorlayout}). 
\subsection{Performance of the TOF-iCT demonstrator}
\subsubsection{Energy and position dependence of the TOF}\label{sec:tofmeasmeth}\mbox{}\\
In order to assess the homogeneity of the TOF measurement for the entire active area of the TOF-iCT demonstrator, the position dependence of the TOF was investigated using different proton beams with energies ranging from $\SI{83}{MeV}$ to $\SI{800}{MeV}$. This was done by projecting the TOF of every particle (equation (\ref{eq:tofinscanner})) onto the last 4D-tracking layer by combining all 86 LGAD channels per sensor to $86\times 86$ pixels. For each pixel, the TOF distribution was recorded and the corresponding most probable value and standard deviation were estimated using a Gaussian fit. The resulting TOF values were then compared to the theoretical TOF through vacuum for the same flight path length and primary beam energy to assess the accuracy of the TOF measurement.
\subsubsection{Energy and position dependence of the intrinsic time resolution}\mbox{}\\
Using the same data set as obtained in section \ref{sec:tofmeasmeth}, the intrinsic time resolution per LGAD channel was estimated for all of the employed beam energies. This was done by calculating the time difference 
\begin{equation}\label{eq:tdifflayer}
    \mathrm{T}_\mathrm{Diff}(i,j)=\mathrm{ToA}_\mathrm{corr}(i_\mathrm{ref},k\neq j)-\mathrm{ToA}_\mathrm{corr}(i,j)
\end{equation} 
between every channel $i$ in $\mathrm{LGAD}_\mathrm{j}$ and a reference channel $i_\mathrm{ref}$ on $\mathrm{LGAD}_\mathrm{k\neq j}$ of the same 4D-tracking layer (figure \ref{fig:demonstratorlayout}). 
A Gaussian was then fitted to the resulting $\mathrm{T}_\mathrm{Diff}(i,j)$ distributions to estimate the corresponding standard deviations $\sigma_{\mathrm{T}_\mathrm{Diff}}(i,j)$. Finally, to approximate the intrinsic time resolution for every LGAD channel ($\sigma_\mathrm{ToA}(i,j)$), the obtained $\sigma_{\mathrm{T}_\mathrm{Diff}}(i,j)$ was divided by $\sqrt{2}$ since $\sigma_{\mathrm{T}_\mathrm{Diff}(i,j)}\approx \sqrt{2} \cdot \sigma_\mathrm{ToA}(i,j)$, which can be derived from a Gaussian error propagation of equation (\ref{eq:tdifflayer}) and assuming a similar time-resolution in each LGAD ($\sigma_\mathrm{ToA}(i,j)\approx \sigma_\mathrm{ToA}$) for a fixed beam energy.  
  
\subsection{Sandwich TOF proton radiography at MedAustron}
\begin{figure}[ht]
	\centering
 \begin{subfigure}[b]{0.48\textwidth}
		\centering
		\includegraphics[height=7cm]{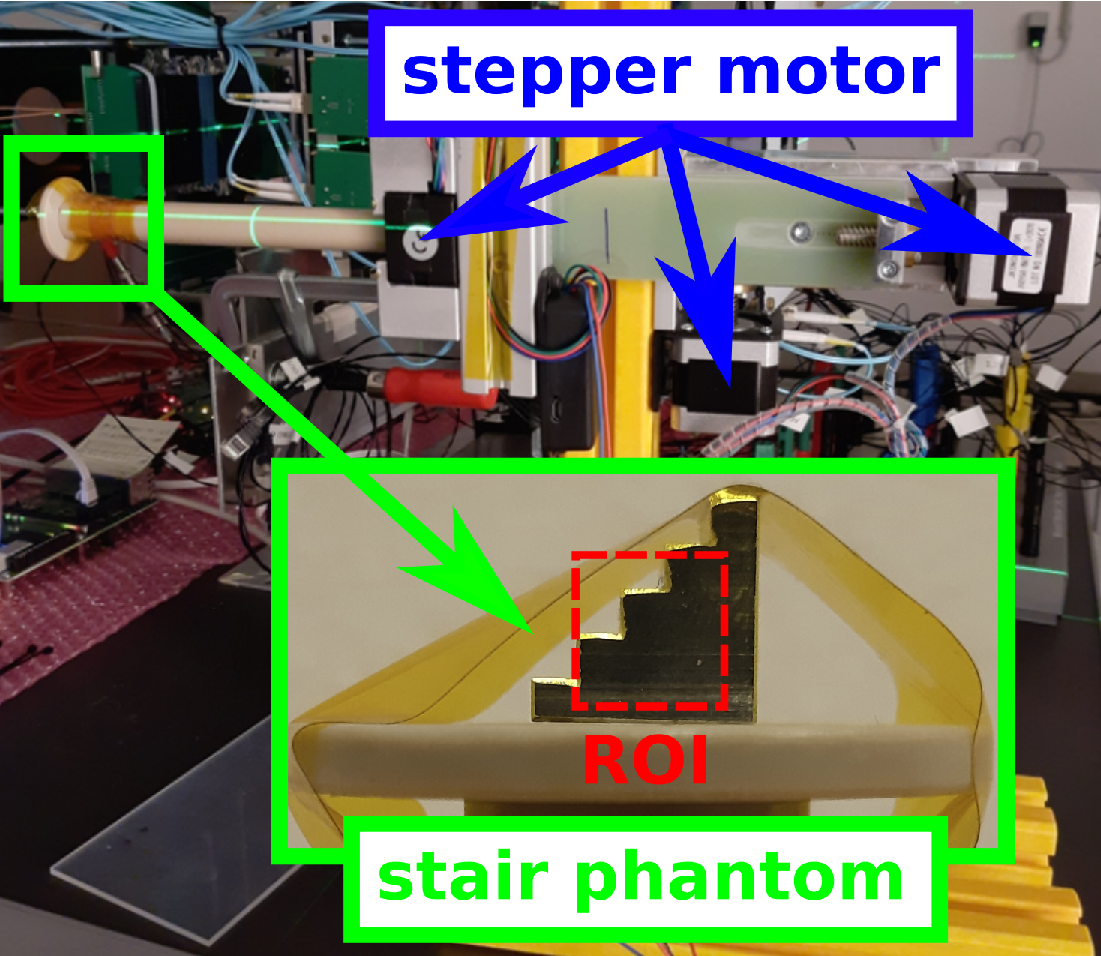}
		\caption{Al stair phantom mounted on a rotating table. The imaged part of the phantom is highlighted as a red region of interest (ROI). }
		\label{fig:stairphan}
	\end{subfigure}
 \quad
	\begin{subfigure}[b]{0.48\textwidth}
		\centering
		\includegraphics[height=7cm]{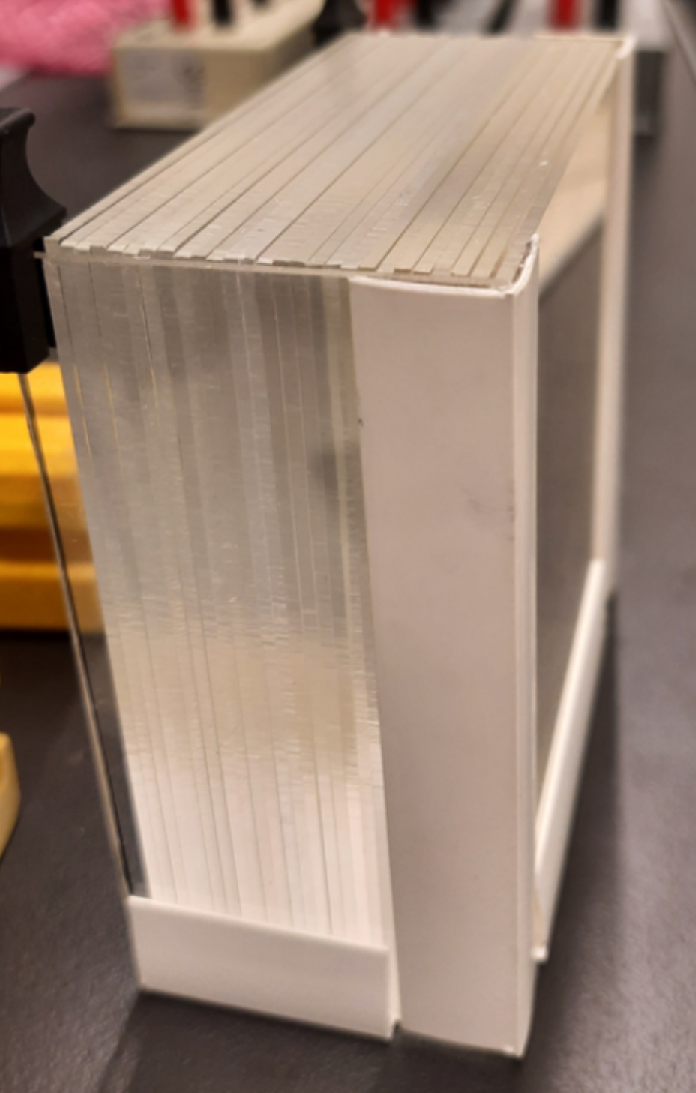}
		\caption{WET calibration phantom consisting of $\SI{10}{}\times \SI{10}{cm^2}$ PMMA slabs with a thickness of $\SI{1.66}{mm}$ each.}
		\label{fig:calibphan}
	\end{subfigure}
	\caption{Calibration phantom and object to be imaged.}
	
\end{figure}

After calibrating and testing the performance of the TOF-iCT demonstrator, two Sandwich TOF pRad images of a $\SI{1}{cm^3}$ aluminium stair phantom (figure \ref{fig:stairphan}) were recorded at MedAustron using $\SI{83}{MeV}$ and $\SI{100.4}{MeV}$ protons. A beam with a reduced particle rate \cite{ulrich-pur_commissioning_2021} close to the maximum trigger rate of the TRB DAQ system ($\approx\SI{e5}{p/s}$) was chosen to reduce the probability of multiple particle hits per LGAD for every triggered event. However, this also meant that the full potential of the rate capability of LGADs ($\approx \SI{e8}{p/s/cm^2}$) was not exploited for this experiment. The main reason for that was to obtain clean 4D particle tracks and avoid any ambiguities caused by e.g. detection inefficiencies in one of the LGADs, which would then require a more sophisticated 4D-tracking algorithm. The latter was out of the scope, as the main focus of this study was the first experimental realisation of Sandwich TOF
pRad, for which a clean particle track and, therefore, simpler 4D-tracking methods were preferred.
\subsubsection{WET calibration}
\label{sec:wetcalibmethsetup}\mbox{}\\
First, a calibration curve according to section \ref{sec:wetcalibmeth} was recorded for each beam energy. This involved irradiating $\SI{1.65}{mm}$ thick PMMA slabs (figure \ref{fig:calibphan}) with an RSP of $\SI{1.012}{}$ and measuring the corresponding TOF per pixel inside the TOF-iCT demonstrator as detailed in section \ref{sec:tofmeasmeth}. Then, the median over those TOF per pixel values was calculated and used to determine the calibration parameters $a_i(E_0)$ (see equation (\ref{eq:WETcalibmodel})).
\begin{figure}[ht]
\centering
	\begin{subfigure}[b]{0.9\textwidth}
		\centering
		\includegraphics[width=0.9\textwidth]{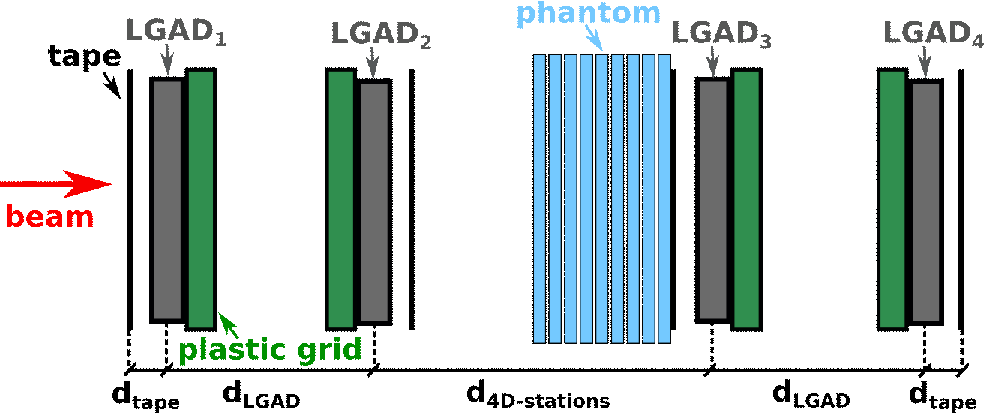}
		\caption{Schematic representation of the WET calibration setup at MedAustron.  }
		\label{fig:calibphansetup}
	\end{subfigure}
	\quad
	\begin{subfigure}[b]{0.9\textwidth}
		\centering
		\includegraphics[width=0.9\textwidth]{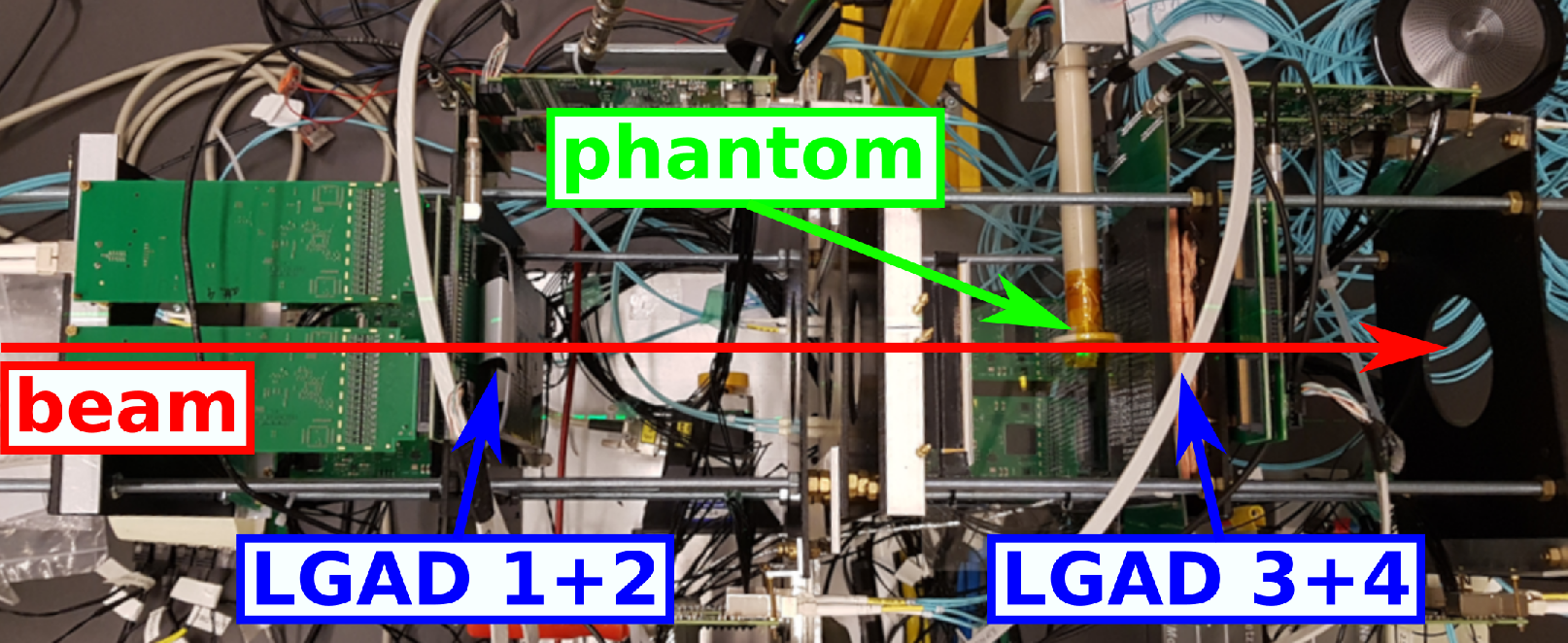}
		\caption{Image of the TOF-pCT setup at MedAustron. The Al stair phantom was mounted on a rotating table and placed in front of the second 4D-tracking station marked as LGAD ``3+4''.}
		\label{fig:stairphansetup}
	\end{subfigure}
\vspace{-0.3cm}
\caption{Sandwich TOF-iCT demonstrator system consisting of 4 single-sided LGAD strip sensors. Only every second channel of LGAD3 was connected to the readout electronics.}
\label{fig:setupsketch}
\end{figure}
\noindent As shown in figure \ref{fig:calibphansetup}, the calibration phantom was mounted in front of the third LGAD. It was placed adjacent to the light-tight enclosure covering the LGAD to minimize the energy loss in air between the phantom and the last 4D-tracking station.
\subsubsection{TOF-based proton radiography}\label{sec:pCRmeth}\mbox{}\\
After determining the WET calibration curves, the pRads of the aluminium stair phantom were recorded. For that purpose, the phantom was mounted on a rotating table and placed $\SI{1.9}{cm}$ in front of the third LGAD (figure \ref{fig:stairphansetup}). However, given that each LGAD consisted of 86 strips with a strip pitch of $\SI{100}{\micro m}$, only a partial section of the phantom could be imaged. The specific area imaged is delineated in figure \ref{fig:stairphan} by the red region of interest (ROI).
\\\\
In order to acquire the pRad image for each beam energy, the TOF through the scanner was measured and projected onto the last 4D-tracking layer as outlined in section \ref{sec:tofmeasmeth}. A pixel size of $\SI{2}{}\times \SI{2}{}$ strips was selected for the pRad images, which is equivalent to an area of $\SI{200}{\micro m}\times\SI{200}{\micro m}$ per pixel. For each pixel, the MPV of the TOF, i.e. $\mathrm{TOF}_\mathrm{MPV}(x,y)$, was obtained by applying a Gaussian fit to the resulting TOF distributions. Using the WET calibration curves as determined in section \ref{sec:wetcalibmethsetup}, the collected $\mathrm{TOF}_\mathrm{MPV}(x,y)$ values were converted into the corresponding WET. 
\subsubsection{Image quality analysis}\label{sec:imageanalysismeth}\mbox{}\\
To assess the pRad image quality with respect to the WET accuracy, the WET values were collected inside the centre of the aluminium stair phantom using a square-shaped ROI with a size of $\SI{24}{}\times\SI{24}{}$ pixel. Then, the relative WET error was calculated according to
\begin{equation}
    \label{eq:wetaccuracy}
    \epsilon_\mathrm{WET}=\frac{\mathrm{WET}_\mathrm{median}-\mathrm{WET}_\mathrm{theo}}{\mathrm{WET}_\mathrm{theo}}.
\end{equation}
Here, $\mathrm{WET}_\mathrm{median}$ represents the measured median WET per pixel inside the ROI and $\mathrm{WET}_\mathrm{theo}$ the theoretical WET for a $\SI{1}{cm}$ thick aluminium layer, which, according to the NIST PSTAR database \cite{berger2017stopping}, is $\SI{20.968}{mm}$.\\\\
In addition to the WET accuracy, the WET resolution was estimated. For that purpose, the quartile coefficient of dispersion ($\mathrm{QCOD}_\mathrm{WET}$) was calculated inside the previously specified ROI using
\begin{equation}
    \label{eq:qcod}
    \mathrm{QCOD}_\mathrm{WET}=\frac{\mathrm{Q}_\mathrm{3,WET}-\mathrm{Q}_\mathrm{1,WET}}{\mathrm{Q}_\mathrm{3,WET}+\mathrm{Q}_\mathrm{1,WET}},
\end{equation}
with $\mathrm{Q}_\mathrm{1,WET}$ and $\mathrm{Q}_\mathrm{3,WET}$ as the first and third quartiles and $\mathrm{Q}_\mathrm{3,WET}-\mathrm{Q}_\mathrm{1,WET}$ as the interquartile range ($\mathrm{IQR_\mathrm{WET}}$) of the corresponding WET distribution.
\section{Results}
\subsection{Sensor calibration and 4D hit reconstruction}
This section presents the main results of the sensor calibration steps for the LGADs and 4D hit reconstruction methods as described in sections \ref{sec:sensorcalibmeth} and \ref{sec:4dtrackandclustmeth}.
\subsubsection{Time-over-threshold rescaling and 4D clustering}\label{sec:restotrescaling4dclust}
\begin{figure}[ht]
	\centering
	\begin{subfigure}[b]{0.48\textwidth}
		\centering
		\includegraphics[width=0.99
		\textwidth]{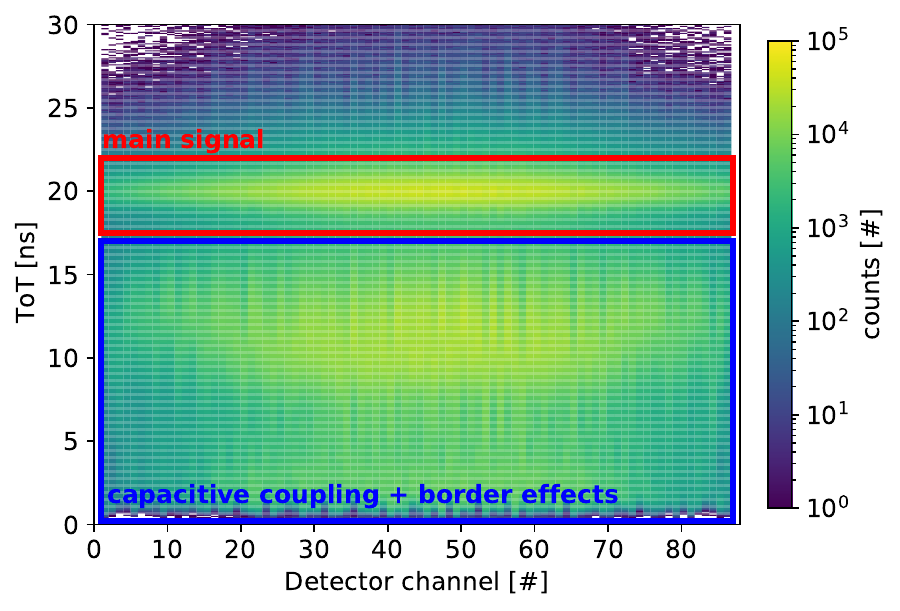}
		\vspace{-0.3cm}
		\caption{Re-scaled ToT inside a single LGAD showing the main signal peak and the hits in the neighbouring strips (capacitive coupling).}
		\label{fig:800capacoupl}
	\end{subfigure}
	\quad
	\begin{subfigure}[b]{0.48\textwidth}
		\centering
		\includegraphics[width=0.99\textwidth]{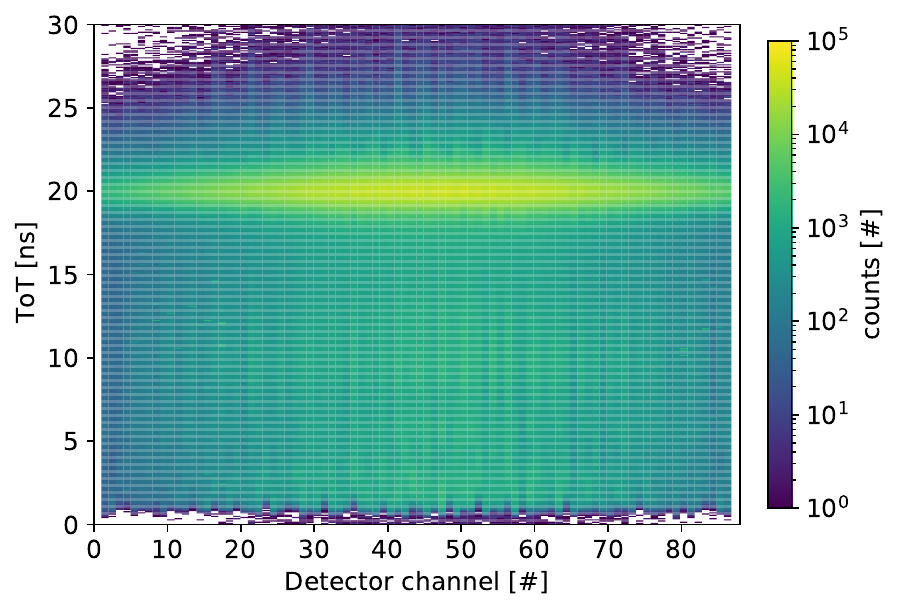}
		\vspace{-0.3cm}
		\caption{ToT of the main signal peak defined by the local maxima inside the obtained 4D clusters.}
		\label{fig:800cluster}
	\end{subfigure}
	
	\caption{Effect of cluster finder algorithm shown for $\SI{800}{MeV}$ protons measured inside the third LGAD. }
\end{figure}

\mbox{}\\
Figure \ref{fig:800capacoupl} shows a normalized ToT distribution within a single LGAD at $\SI{800}{MeV}$. As outlined in section \ref{sec:totnormmeth}, the primary signal peak (marked in red) was shifted to $\SI{20}{ns}$ for each beam energy, ensuring a consistent response across all LGAD channels. Besides this main signal peak, also secondary peaks are present at lower ToT values (highlighted in blue), which originate from capacitively-coupled hits in the strips adjacent to the main hit position.
\noindent In order to remove those capacitive-coupled signals, the 4D-cluster finder algorithm (section \ref{sec:4dclustgmeth}) was applied to the recorded data set. Depending on the used beam energy, the mean cluster size, i.e. the number of fired strips per 4D cluster, ranged between three ($\SI{800}{MeV}$) and five ($\SI{83}{MeV}$) strips. The main reason is that, for lower beam energies, the signal increases, leading to signals above threshold even at strips further away from the actual hit strip.\\
Once the clusters were identified, the strips with the highest ToT values were used as an estimate for the true particle hit position. The corresponding ToT distribution is depicted in figure \ref{fig:800cluster}. Now, only the main signal peak can be recognized. However, a low-ToT background with much lower statistics than the main signal is still visible. As described in \citeasnoun{Pietraszko2020}, this low-ToT background stems from border effects at certain regions inside the sensor, which show a lower gain than the nominal one, e.g. between two strips or at the border of the sensor. 
\begin{figure}[ht]
	\centering
 \begin{subfigure}[b]{0.48\textwidth}
		\centering
		\includegraphics[width=0.99
		\textwidth]{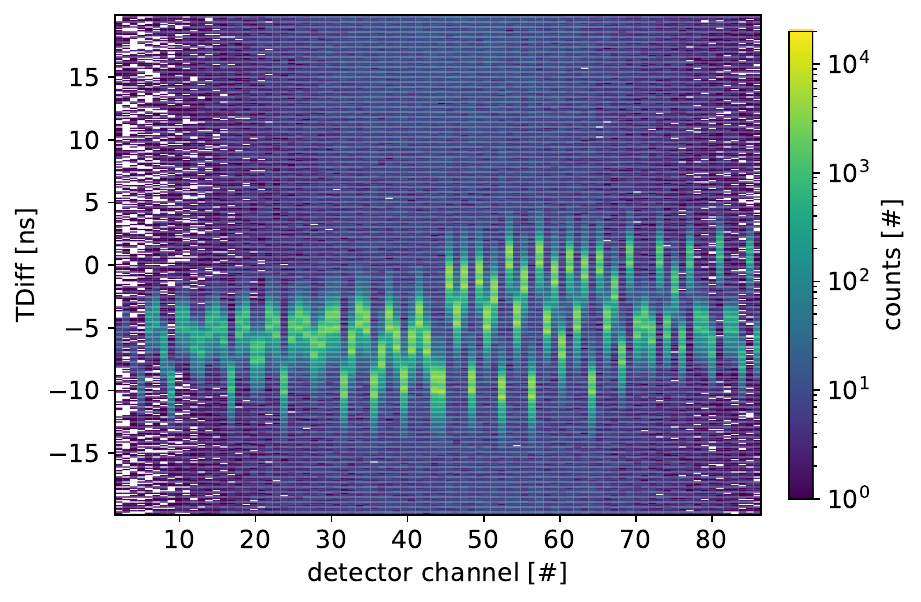}
		\vspace{-0.3cm}
		\caption{Raw time difference spectrum for a single LGAD sensor with channel-dependent offsets. }
		\label{fig:800det3offset}
	\end{subfigure}
	\quad
	\begin{subfigure}[b]{0.48\textwidth}
		\centering
		\includegraphics[width=0.99\textwidth]{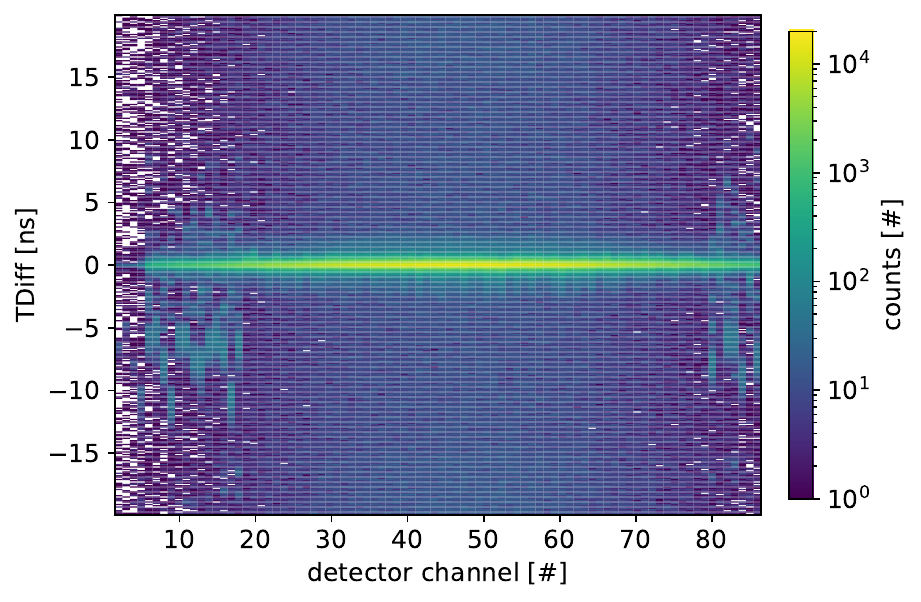}
		\vspace{-0.3cm}
		\caption{Time-walk and offset corrected data for a single LGAD sensor.}
		\label{fig:800det3offsetpostcalib}
	\end{subfigure}
	\begin{subfigure}[b]{0.48\textwidth}
		\centering
		\includegraphics[width=0.99
		\textwidth]{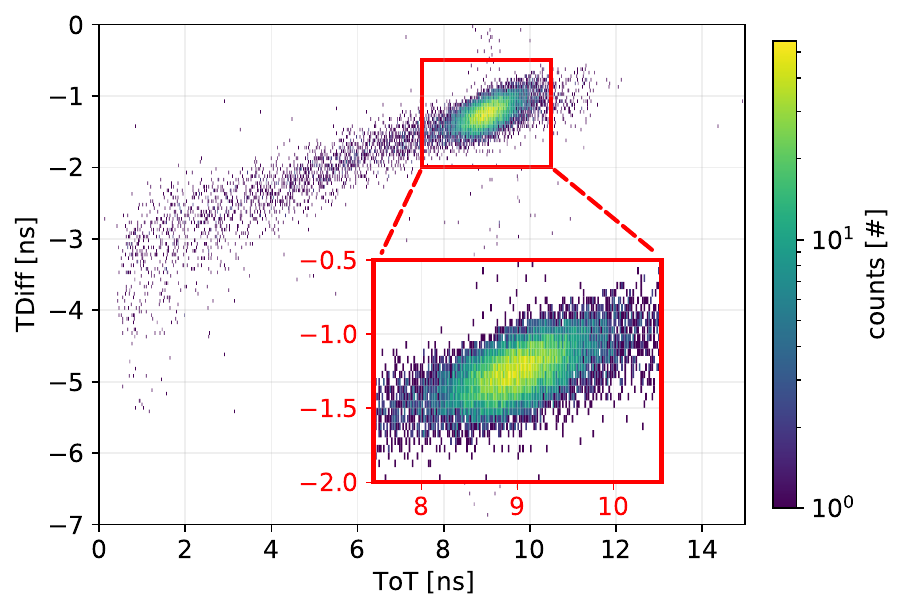}
		\vspace{-0.3cm}
		\caption{ToA distribution for a single LGAD channel. The time-walk effect is clearly visible in the zoomed-in signal peak distribution (highlighted in red).}
		\label{fig:800praetwalk}
	\end{subfigure}
	\quad
	\begin{subfigure}[b]{0.48\textwidth}
		\centering
		\includegraphics[width=0.99\textwidth]{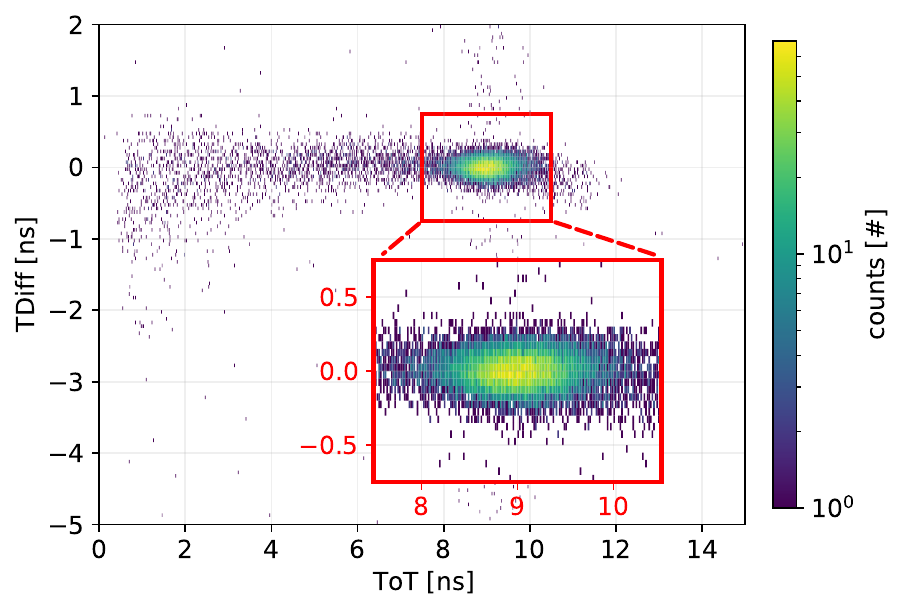}
		\vspace{-0.3cm}
		\caption{Time-walk corrected ToA distribution for a single LGAD channel. The zoomed-in signal peak distribution (highlighted in red) shows no ToT-dependence anymore.}
		\label{fig:800posttwalk}
	\end{subfigure}
	
	\caption{Time-walk and offset correction for $\SI{800}{MeV}$ protons.}
\end{figure}
\subsubsection{Time-walk and offset calibration}\label{sec:resrestimewalk}\mbox{}\\
 Without any time-walk and offset correction, the time difference between the individual LGAD sensors per 4D-tracking layer shows a strong channel dependence due to different propagation times and time-walk properties in each channel (figure \ref{fig:800det3offset}). However, after applying the time-walk and offset corrections as outlined in section \ref{sec:timewalkmeth}, a uniform response across the whole sensor for the given beam energy could be achieved. The resulting mean time difference spectrum was then centred at $\SI{0}{ns}$ and showed a reduced variability in each channel (figure \ref{fig:800det3offsetpostcalib}). 
 The latter is caused by the time-walk calibration, as the time-walk effect widens the time difference distribution due to its signal amplitude dependence of the ToA. To demonstrate this time-walk effect, figure \ref{fig:800praetwalk} depicts an example time difference vs ToT distribution for a single LGAD channel with respect to a reference channel on the second LGAD of the same 4D-tracking layer.  As mentioned in section \ref{sec:timewalkmeth}, a ToT cut was then applied on the reference channel to guarantee that the hits on the first LGAD were only correlated once with the true particle hit on the reference channel of the second LGAD. Although those ToT cuts for the reference channel were applied on the normalized ToT values, the time-walk calibration was done using the raw, non-normalized ToT values ($\mathrm{TOT}_\mathrm{raw}$), as this provided a more accurate correction of the time-walk effect. Thus, the ToT values depicted in figure \ref{fig:800praetwalk} also represent the $\mathrm{TOT}_\mathrm{raw}$. In this graph, the signal amplitude-dependence of the ToA, which is in the order of $\approx\SI{}{ns}$, is clearly visible. To mitigate this time-walk effect, we conducted a time-walk calibration as outlined in section \ref{sec:timewalkmeth}. The resulting time difference distribution for the calibrated data is depicted in figure \ref{fig:800posttwalk}. No further ToT-dependence on the ToA can be recognized and the main TDiff signal peak is now centred around zero.
 \subsection{4D-tracking}\label{sec:res4dtracking}
 After calibrating the LGAD sensors, the TOF-iCT demonstrator was aligned according to section \ref{sec:alignmentmeth}. The resulting offset parameters were $\hat{r}_x\approx\SI{1.03}{mm}$ for the x-direction and $\hat{r}_y\approx\SI{0.59}{mm}$ for the y-direction. Using the aligned and calibrated TOF-iCT demonstrator, the TOF through the demonstrator was recorded for several beam energies ranging between $\SI{83}{MeV}$ and $\SI{800}{MeV}$ to study the energy and position dependence of the 4D-track reconstruction.
 \subsubsection{Energy and position dependence of the 4D-track reconstruction}
\begin{figure}[ht]
	\centering
	\begin{subfigure}[b]{0.48\textwidth}
		\centering
		\includegraphics[width=0.915
		\textwidth]{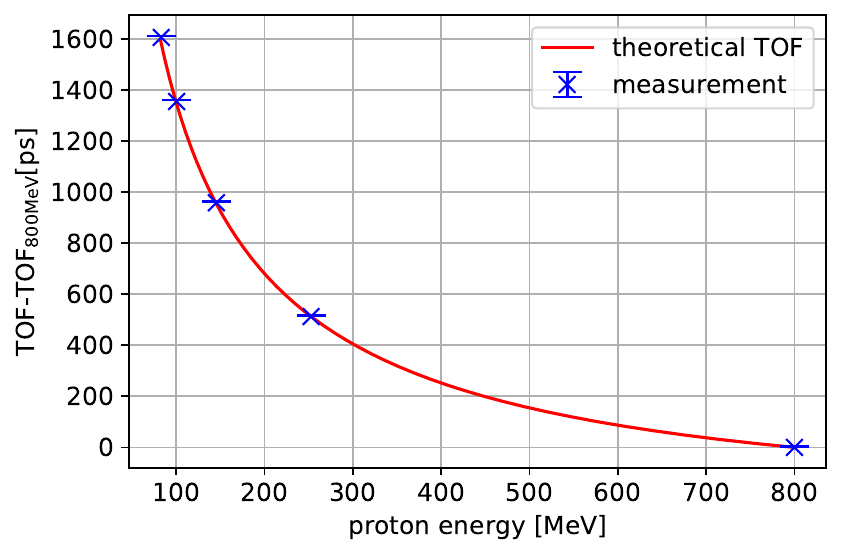}
		\vspace{-0.3cm}
		\caption{Measured TOF in air for different energies compared to the theoretical TOF values.}
		\label{fig:tofinair}
	\end{subfigure}
	\quad
	\begin{subfigure}[b]{0.48\textwidth}
		\centering
		\includegraphics[width=0.983\textwidth]{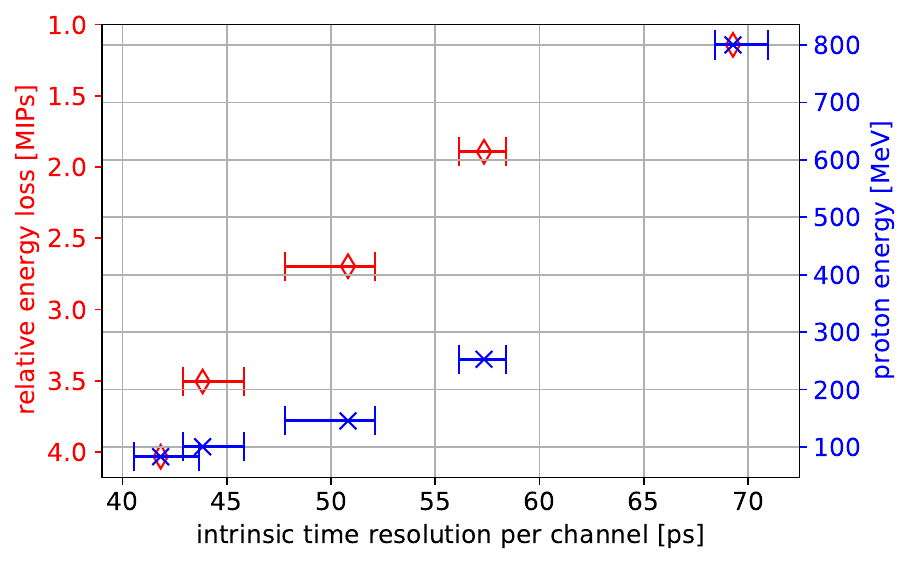}
		\vspace{-0.3cm}
		\caption{Timing precision for various beam energies measured in the first LGAD.}
		\label{fig:teresenergy}
	\end{subfigure}
	
	\caption{Energy and position dependence of the TOF and time-resolution.}
\end{figure}

\mbox{}\\
 For each of the used beam energies, the median TOF per pixel and interquartile range were calculated. Then, the median TOF per pixel was compared to the calculated TOF in vacuum after subtracting the corresponding TOF values at $\SI{800}{MeV}$ (figure \ref{fig:tofinair}). 
 \noindent Setting the TOF at $\SI{800}{MeV}$ to zero enabled the correction of any extra offset between the measured and theoretical TOF values, which allowed for a more accurate comparison. While the relative difference between the measured median TOF and TOF in vacuum was $\SI{1.55}{\percent }$ for $\SI{83}{MeV}$, it was lower than $\SI{0.74}{\percent }$ for all other beam energies and reached its lowest value of $\SI{0.004}{\percent }$ at $\SI{252.7}{MeV}$. This decrease at higher beam energies can be explained by the reduced energy loss along the particle path at increased beam energies.\\
Using the same data sets, the intrinsic time resolution per LGAD was calculated according to section \ref{sec:tofmeasmeth}. The resulting time resolution per pixel and corresponding interquartile ranges are depicted in figure \ref{fig:teresenergy}. While the blue data points represent the intrinsic time resolution with respect to the primary beam energy, the red diamond-shaped data points represent the corresponding energy loss in MIPs, i.e. the ratio between the expected energy loss for the used beam energy and the expected energy loss of a MIP. Both of those values were calculated using the NIST PSTAR database \cite{berger2017stopping}. As shown in figure \ref{fig:teresenergy}, the intrinsic time resolution strongly depends on the beam energy and improves with lower beam energies, i.e. higher energy loss inside the detector. While $\SI{800}{MeV}$ yielded a median time resolution of $\SI{69.3}{ps}$, $\SI{83}{MeV}$ showed an intrinsic time resolution of $\SI{41.8}{ps}$.
\subsection{WET calibration}\label{sec:reswetcalib}
\begin{figure}[ht]
\begin{center}
\includegraphics[width=0.5
\textwidth]{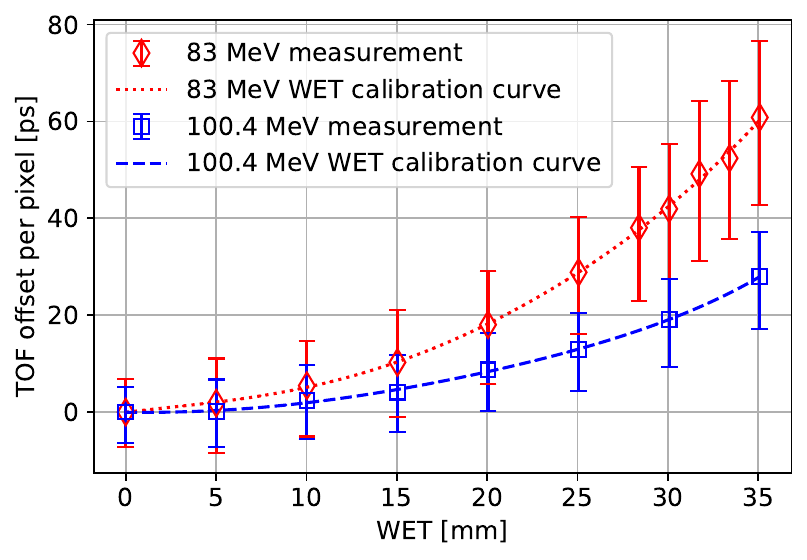}
\end{center}
\vspace{-0.3cm}
\caption{Material and energy-dependent TOF increase for different PMMA absorber thicknesses. The squared and diamond-shaped data points indicate the median TOF increase across all pixels and the error bars denote the corresponding interquartile ranges. A fifth-order polynomial was fitted to the median TOF increase to obtain the WET calibration curve for each primary beam energy (dashed and dotted lines).}
\label{fig:calibcurves}
\end{figure}

Figure \ref{fig:calibcurves} depicts the TOF offset per pixel for different absorber thicknesses and beam energies with respect to the TOF per pixel without any absorber, i.e. $\mathrm{WET}=\SI{0}{mm}$. To highlight the TOF variability between the individual LGAD channels, the median TOF per pixel and corresponding interquartile ranges are given for each data point.  
\noindent While the red diamond-shaped markers denote the median TOF values for $\SI{83}{MeV}$, the blue square-shaped data points represent the median TOF values for $\SI{100.4}{MeV}$. To determine the calibration parameters $a_i(E_0)$ as described in equation (\ref{eq:WETcalibmodel}), a fifth-order polynomial was then fitted to the median TOF values across all pixels. The corresponding WET calibration curves for $\SI{83}{MeV}$ and $\SI{100.4}{MeV}$ are indicated as the dotted and dashed lines, respectively. As expected, the TOF through the absorber increases with higher energy loss inside the phantom, e.g. caused by thicker absorbers or lower beam energies. Similarly, the TOF variability, which is indicated by the error bars representing the interquartile range (IQR), increases with larger WET values. This trend can be attributed to the increased energy straggling inside the phantom for larger phantom thicknesses.


\subsection{TOF-based proton radiography}
\begin{figure}[ht]
	\centering
	\begin{subfigure}[b]{0.48\textwidth}
		\centering
		\includegraphics[width=0.9\textwidth]{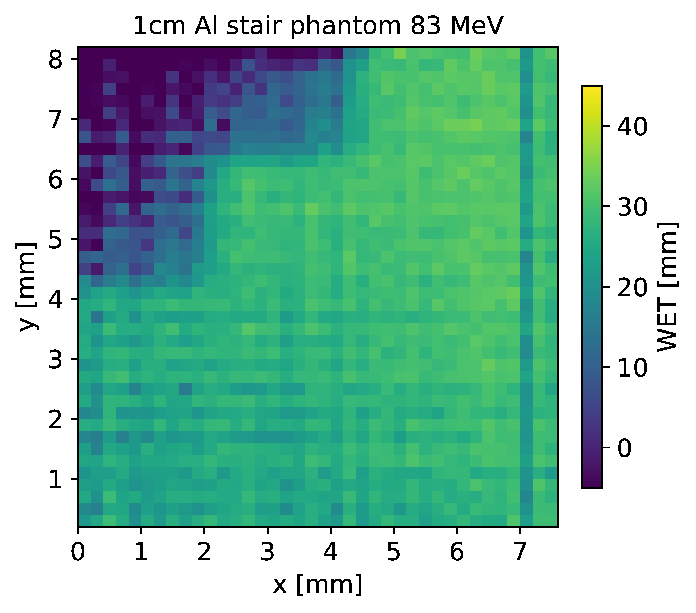}
		\caption{pRad recorded at $\SI{83}{MeV}$.}
		\label{fig:83Mevoffsetperpixel}
	\end{subfigure}
	\begin{subfigure}[b]{0.48\textwidth}
		\centering
		\includegraphics[width=0.9\textwidth]{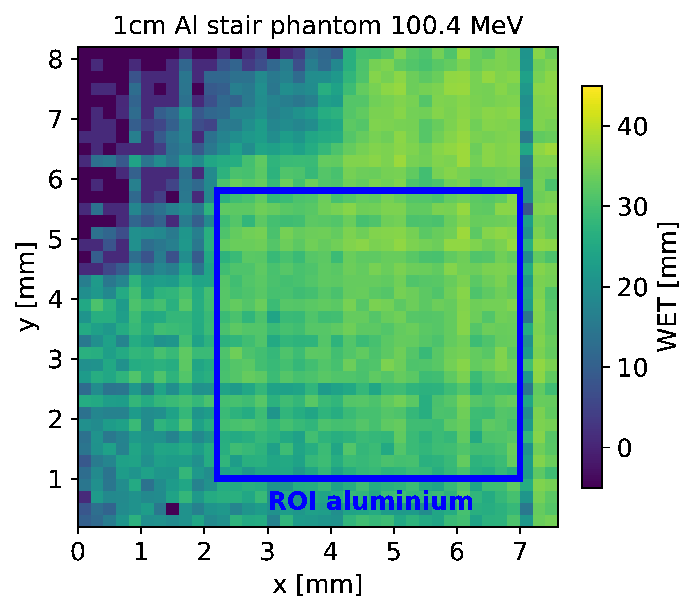}
		\caption{pRad recorded at $\SI{100.4}{MeV}$.}
		\label{fig:100MeVoffsetperpixel}
	\end{subfigure}
	\caption{Obtained pRad images for $\SI{83}{MeV}$ (left) and $\SI{100.4}{MeV}$ (right). The WET values were collected in a squared ROI, which is indicated in blue on the right (ROI aluminium).}
\end{figure}

\begin{figure}[ht]
	\centering
	\begin{subfigure}[b]{0.48\textwidth}
		\centering
		\includegraphics[width=0.99\textwidth]{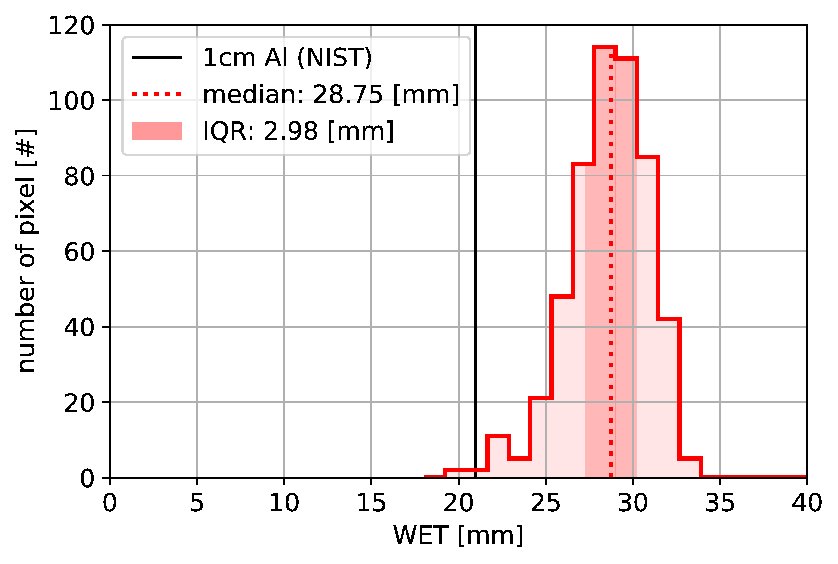}
		\caption{WET measurement for $\SI{83}{MeV}$.}
		\label{fig:83Mevwetdistr}
	\end{subfigure}
	\begin{subfigure}[b]{0.48\textwidth}
		\centering
		\includegraphics[width=0.99\textwidth]{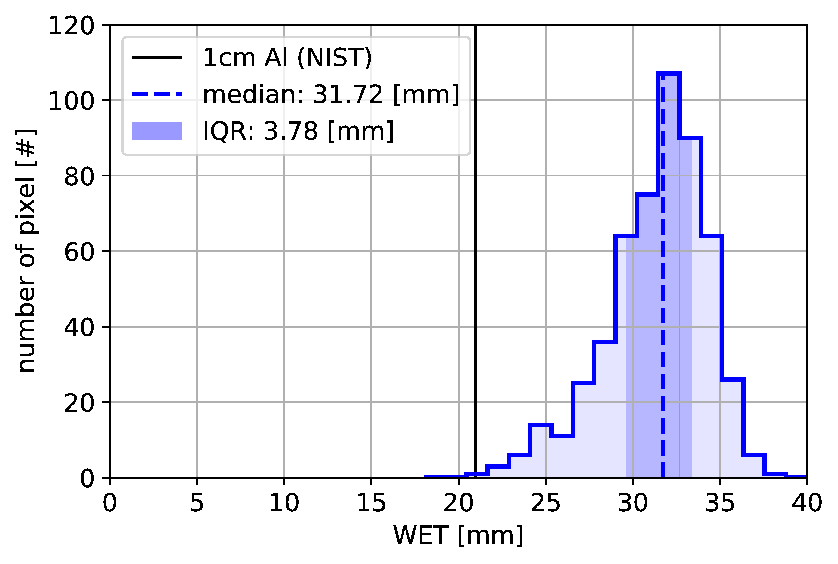}
		\caption{WET measurement for $\SI{100.4}{MeV}$.}
		\label{fig:100MeVwetdistr}
	\end{subfigure}
	\caption{WET distribution inside the ROI and corresponding median and inter-quartile range (IQR). The IQR is also highlighted in dark red and dark blue for $\SI{83}{MeV}$ and $\SI{100.4}{MeV}$, respectively.}
 \label{fig:wetdistrresults}
\end{figure}

The recorded pRad images are shown in figures \ref{fig:83Mevoffsetperpixel} ($\SI{83}{MeV}$) and \ref{fig:100MeVoffsetperpixel} ($\SI{100.4}{MeV}$). In both radiographs, the phantom is clearly visible and the contours of the steps can be recognized. The vertical streak, which appears in both images at $x \approx \SI{7}{mm}$, is an artifact which stems from noisy LGAD channel at this position. 
\mbox{}\\
As outlined in section \ref{sec:imageanalysismeth}, to assess the image quality, the WET values were collected in a squared-shaped ROI, which is indicated in blue in figure \ref{fig:100MeVoffsetperpixel} (ROI aluminium). The resulting WET distributions are given in figures \ref{fig:83Mevwetdistr} ($\SI{83}{MeV}$) and \ref{fig:100MeVwetdistr} ($\SI{100.4}{MeV}$). 

\subsubsection{WET accuracy}\mbox{}\\For each beam energy, the median WET was then calculated, represented by the dotted ($\SI{83}{MeV}$) and dashed ($\SI{100.4}{MeV}$) lines in figure \ref{fig:wetdistrresults}. The determined $\mathrm{WET}_\mathrm{median}$ values were $\SI{28.75}{mm}$ and $\SI{31.72}{mm}$ for $\SI{83}{MeV}$ and $\SI{100.4}{MeV}$, respectively. In order to compare the obtained median WET values to the theoretical WET from the NIST database ($\SI{20.968}{mm}$), the relative WET errors were estimated according to equation (\ref{eq:wetaccuracy}). The results are summarised in table \ref{tab:imageresults} showing a relative WET error of $\epsilon_\mathrm{WET} =\SI{37.09}{\percent}$ for $\SI{83}{MeV}$ and $\epsilon_\mathrm{WET} =\SI{51.12}{\percent}$ for $\SI{100.4}{MeV}$. 
\begin{table}[ht!]
\centering
\begin{tabular}{lcc}
\toprule
\textbf{Parameter} & \textbf{83 MeV} & \textbf{100 MeV} \\ \midrule
$\bm{\mathrm{WET}_\mathrm{theo}}$             &    \multicolumn{2}{c}{$\SI{20.968}{mm}$}     \\ 
$\bm{\mathrm{WET}_\mathrm{median}}$             &    $\SI{28.75}{mm}$             &            $\SI{31.72}{mm}$      \\ 
$\bm{\epsilon_\mathrm{WET}}$    &  $\SI{37.09}{\percent}$                &   $\SI{51.12}{\percent}$               \\ 
$\bm{\mathrm{IQR}_\mathrm{WET}}$ &  $\SI{2.98}{mm}$              &  $\SI{3.78}{mm}$                \\ 
$\bm{\mathrm{QCOD}_\mathrm{WET}}$               &      $\SI{5.19}{\percent}$           &   $\SI{6.00}{\percent}$               \\ 
\bottomrule
\end{tabular}
\caption{Summary of the pRad image quality parameters obtained in the ROI.}
\label{tab:imageresults}
\end{table}
\subsubsection{WET resolution}\mbox{}\\As described in section \ref{sec:imageanalysismeth}, the WET resolution was estimated by calculating $\mathrm{QCOD}_\mathrm{WET}$ according to equation (\ref{eq:qcod}). The corresponding IQR is highlighted as the darker-shaded area in figures \ref{fig:83Mevwetdistr} and \ref{fig:100MeVwetdistr}. While the pRad at $\SI{83}{MeV}$ yielded a $\mathrm{QCOD}_\mathrm{WET}$ of $\SI{5.19}{\percent}$, the pRad at $\SI{100.4}{MeV}$ showed a slightly higher $\mathrm{QCOD}_\mathrm{WET}$ of $\SI{6}{\percent}$ (table \ref{tab:imageresults}). 
\section{Discussion}
Initial experimental efforts to develop a TOF-iCT system based on a TOF-calorimeter with Large Area Picosecond Photon Detectors (LAPPDs) have been undertaken in the past \cite{worstell2019}. However, until now, no experimental TOF-iCT has been recorded, as recent proof-of-principle studies were mainly restricted to Monte Carlo simulations \cite{TOFcalUlrich-Pur_2022,krah_relative_2022,sandwichTOFUlrich-Pur_2023}. This work aims to bridge this gap by presenting the first experimental TOF-based pRad using LGAD strip sensors. While developing a clinically viable iCT system was beyond the scope of this study, we were still able to identify experimental challenges that could arise when constructing and calibrating a TOF-iCT system. Those challenges together with the main results of the TOF-pRad experiments are, therefore, discussed in the following.
\subsection{LGAD performance}
A TOF-iCT demonstrator consisting of four single-sided LGAD strip sensors was successfully built and tested. The results from sections \ref{sec:restotrescaling4dclust} and \ref{sec:resrestimewalk} highlight the importance of a proper offset and time-walk calibration to synchronise the individual LGAD channels and to obtain a uniform response across all sensors (figure \ref{fig:800det3offsetpostcalib}). Without those calibration steps, the material and energy-dependent increase in the TOF could not be resolved. The main reason is that both the offsets between the channels and the time-walk effect are in the order of $\approx \SI{}{ns}$ (figure \ref{fig:800det3offset}), which is in the same order of magnitude as the TOF through the scanner (figure \ref{fig:tofinair}) and two orders of magnitude greater than the measured TOF increase (figure \ref{fig:calibcurves}).\\ After fully calibrating the LGAD sensors, the excellent 4D-tracking properties of LGADs could be demonstrated. Depending on the used beam energy, the median intrinsic time resolution per LGAD channel ranged between $\SI{41.8}{ps}$ and $\SI{69.3}{ps}$. While the best time resolution ($\SI{41.8}{ps}$) was obtained at $\SI{83}{MeV}$, the worst time resolution ($\SI{69.3}{ps}$) was observed at $\SI{800}{MeV}$. This energy dependence of the intrinsic time resolution can be explained by the larger signals at lower beam energies and, therefore, also larger signal-to-noise ratio (SNR), which, as shown in \citeasnoun{Sadrozinski2017}, strongly influences the precision of the time measurement. Since LGADs have been mainly used for high energy physics experiments with MIP-like particles, the corresponding intrinsic time resolutions are also usually measured for MIPs \cite{Giacomini2023}. However, as shown in figure \ref{fig:teresenergy}, protons with clinical energies ($<\SI{300}{MeV}$) typically result in more than three times larger signals than MIPs, while even larger signals can be expected for heavier ions, e.g. helium or carbon ions. When looking at figure \ref{fig:teresenergy}, is also apparent that the intrinsic time resolution does not necessarily increase linearly with the signal strength, as it also depends on other factors, e.g. the used LGAD technology, sensor geometry and front-end electronics \cite{Sadrozinski2017}. Thus, when simulating TOF-iCT systems, the actual energy dependence of the time resolution should be taken into account to better approximate the true response of the LGADs to the given experimental conditions. This relation can either be obtained via prior measurements or by performing a more thorough simulation of the signal generation inside the employed LGAD systems e.g. via simulations tools as described in \citeasnoun{Spannagel2022} or \citeasnoun{Cenna2015}.
\subsection{WET calibration}\label{sec:discwetcalib}
A WET calibration was performed according to section \ref{sec:wetcalibmethsetup}. In contrast to the proof-of-concept simulation study described in \citeasnoun{sandwichTOFUlrich-Pur_2023}, where the WET calibration phantom was centred between the innermost LGAD sensors, the WET phantom of this study was mounted directly in front of the third LGAD. This was done to minimize the contribution of the energy-dependent TOF in air downstream of the phantom  (equation (\ref{eq:tof})) to the total measured TOF increase, which, ideally, should only reflect the TOF increase caused by the energy loss inside the scanned object. However, in a clinical scenario, clearances greater than $\SI{10}{cm}$ between the scanner and the patient are typically used since the detectors cannot be placed directly in front of the patient. Thus, an additional TOF offset, depending on the residual beam energy, can be expected, which, currently, is not taken into account in the employed WET calibration method. To reduce this effect, a more sophisticated WET calibration procedure should be considered, which, however, was out of the scope, since this study mainly focused on the first experimental realisation of a TOF-based pRad. 
\subsection{Proton radiography}
As shown in figures \ref{fig:83Mevoffsetperpixel} and \ref{fig:100MeVoffsetperpixel}, two Sandwich TOF-pRads of an aluminium stair phantom could be successfully created after measuring the material-energy dependent TOF increase and mapping it to the corresponding WET. The phantom is clearly visible in both pRads, which were recorded at $\SI{83}{MeV}$ and $\SI{100.4}{MeV}$. However, as outlined in table \ref{tab:imageresults}, the obtained WET values strongly deviated from the expected values. One reason for this discrepancy could be that the aluminium stair phantom could not be mounted directly in front of the third LGAD as it was done for the calibration phantom (section \ref{sec:discwetcalib}). 
Consequently, an additional offset to the measured TOF increase can be expected, caused by the energy-dependent TOF in air between the exit point of the phantom and the third LGAD. This also becomes apparent when looking at equation (\ref{eq:tof}). For the same flight distance, i.e. in this case for the flight pathlength in air downstream of the phantom, the TOF increases with lower beam energy. Since the residual beam energy $E_0-\Delta E$ is always lower than the primary beam energy $E_0$, the TOF for this distance will always be larger if $\Delta E>0$. Thus, the measured TOF increase for each pCR scan includes not only the TOF increase caused by the energy loss inside the phantom but also an additional contribution from the energy-dependent TOF in air downstream of the phantom, which could explain the overestimated WET values. This systematic error is particularly more significant for this small phantom size given the steepness of the calibration curves at the expected WET values (figure \ref{fig:calibcurves}). At $\SI{100.4}{MeV}$, this effect is even more pronounced, as smaller deviations in the measured TOF increase lead to a larger WET difference when compared to $\SI{83}{MeV}$. This increased WET at higher beam energies is also reflected in the WET error, which was always positive and larger for $\SI{100.4}{MeV}$.  \\Another factor, which influences the accuracy of the pRad measurement, is the particle path estimation model. While the authors of \citeasnoun{sandwichTOFUlrich-Pur_2023} used the well-established most-likely-path (MLP) formalism \cite{Schulte2008} in their proof-of-principle simulation study, this work was restricted to a SL approximation due to the limited number of available 4D-tracking layers. Still, the Sandwich TOF-pRads could be recorded using only two 4D-tracking layers. Since an upgrade of the current TOF-iCT demonstrator, including an increased number of 4D-tracking layers, is currently in development, more accurate path estimation models, e.g. the MLP formalism, can be used in the future. This will also include the application of the standard $3$$\sigma$ cuts on the scattering angles to filter large-angle scattering events \cite{Schulte2008}, which, for the aforementioned reasons, could not be applied in this study.
\subsection{TOF-iCT demonstrator}
As mentioned in section \ref{sec:pCRmeth}, the full rate capability of the employed LGAD-based TOF-iCT system was not exploited since the main focus was to guarantee clear 4D particle tracks by minimizing tracking ambiguities caused by multiple hits per LGAD in each triggered event. Thus, the beam rate was set close to the maximum trigger rate of the TRB3 system, which is in the order of $\SI{100}{kHz}$. However, due to additional tracking inefficiencies caused by e.g. limitations of the current trigger logic, the actual measured number of particle tracks was also reduced. The main reason was that an event was triggered whenever any hit inside the third LGAD was observed, which could stem from both a true particle hit or also noise event. To guarantee that only events will be recorded once a particle has traversed all four detectors, an adapted and upgraded trigger logic should be implemented. Furthermore, an improved 4D-track estimation algorithm should be developed to better deal with tracking ambiguities, which will allow for much higher particle rates ($\mathcal{O}(\SI{e8}{p/s/cm^2})$) as demonstrated in \citeasnoun{kruger_lgad_2022}.
\subsection{Demands for a clinical TOF-iCT system}
The current TOF-iCT demonstrator is able to perform first proof-of-principle measurements of small objects, which is essential to gain a better understanding and improve this novel imaging modality. However, to be able to apply Sandwich TOF-iCT to a more clinically relevant scenario, e.g. imaging of a human head phantom, a large-area TOF-iCT system has to be developed. Currently, one of the most limiting factors with respect to the scanner size is the discrete nature of the analogue FEE, which limits the number of possible readout channels per detector area. Thus, a dedicated application-specific integrated circuit (ASIC), able to handle the readout of tens of readout channels without distorting the quality of the time measurement has to be developed. Presently, the most advanced ASICs for the readout of large-area LGAD systems are the ETROC \cite{etroc} and ALTIROC \cite{altiroc} ASICs from the ATLAS and CMS collaborations, which are used to read out active detector areas in the order of $\SI{}{m^2}$ \cite{atlas_technical_2018,cms_mip_2019}. Both ASICs contain the full pre-amplification and digitization chain for an LGAD while keeping the power consumption as low as possible. However, the design of those ASICs was optimized for the experimental conditions at LHC, in particular the bunch-crossing frequency of $\SI{40}{MHz}$. The ETROC e.g. uses the first half of the $\SI{40}{MHz}$ cycle to record all particles, while the second half is used for data transfer, as no particle interaction can be expected during that period. However, medical accelerators usually deliver the beam in spills with a duration in the order of several seconds. Thus, using an ASIC with the aforementioned logic would lead to a dead time of $\SI{50}{\percent}$ and, therefore, would require twice the dose to complete the image acquisition. Consequently, a dedicated ASIC tailored to the needs of an ion imaging system and experimental conditions of a medical particle accelerator has to be developed.
\\It also has to be mentioned that both the ETROC and ALTIROC were designed for the readout of LGAD pixel sensors, which, for $n$ pixels per spatial coordinate, require the readout of $n^2$ channels and thus lead to higher demands and also costs for the readout electronics. To cover the same detector area with LGAD strip sensors, only $2n$ channels would need to be read out, which would reduce the cost of the readout electronics. If single-sided and not double-sided LGAD strip sensors are used \cite{Bisht2022}, the number of LGAD sensors per 4D-tracking layer will double. While this would diminish the cost advantage over LGAD pixel sensors, at the same time, the time precision of the ToA measurement will improve, as the ToA is determined twice per 4D-tracking layer.\\
Independent of the used detector technology and readout electronics, a dedicated low-mass detector module has to be designed, as the size of the individual LGAD sensors is limited to a few $\SI{}{cm^2}$. Since the final 4D-tracking layer will then comprise of several such LGAD modules, the mass of the readout electronics should be kept low as possible to reduce the material budget of the final 4D-tracking layer and to keep its response as uniform as possible.

\section{Conclusion}
The purpose of this work was to realise the first experimental TOF-based ion imaging experiment by recording a Sandwich TOF-pRad with LGAD strip sensors. While developing a clinically viable iCT system was out of the scope, we could still successfully measure two Sandwich TOF-pRads of a small $\SI{1}{cm}$ thick aluminium stair phantom at MedAustron using $\SI{83}{MeV}$ and $\SI{100.4}{MeV}$ protons. However, the measured WET strongly deviated from the expected values, i.e. $\SI{37.09}{\percent}$ for $\SI{83}{MeV}$ and $\SI{51.12}{\percent}$ for $\SI{100.4}{MeV}$, which was mainly attributed to the simplified WET calibration model employed in this study. This WET calibration model as well as the used 4D-tracking algorithms should be improved in the future, which, however, was not the main focus of this work. Furthermore, experimental challenges and potential improvements could be identified, in particular the necessity for a large-area 4D-tracking system tailored to the experimental conditions of a medical particle accelerator. This also includes the development of a dedicated ASIC and low-mass LGAD module able to read out hundreds of detector channels.
Nevertheless, this study was still an important first step to further advance TOF-iCT, in particular Sandwich TOF-iCT, which once implemented, could provide a more compact and easier-to-integrate solution for iCT.

\section*{Acknowledgements}
This research was funded in whole or in part by the Austrian Science Fund (FWF) Erwin-Schrödinger grant J 4762-N. 

\section*{Data Availability Statement} The data supporting this study are currently not published. However, the data of any Author Accepted Manuscript (AAM) version arising from this submission will be made publicly available on the \url{https://www.hepdata.net} server.

\section*{References}

\bibliography{sandwichtof}
\end{document}